\documentclass[fleqn,usenatbib]{mnras}

\usepackage[T1]{fontenc}

\DeclareRobustCommand{\VAN}[3]{#2}
\let\VANthebibliography\thebibliography
\def\thebibliography{\DeclareRobustCommand{\VAN}[3]{##3}\VANthebibliography}

\usepackage{graphicx}	
\usepackage{amsmath}	
\usepackage{amssymb}	

\usepackage{color}

\usepackage{url}

\title[Doubly eclipsing quadruple system BG Ind]{BG Ind: the nearest doubly eclipsing, compact hierarchical quadruple system}

\author[Borkovits et al.]{
T. Borkovits$^{1,2,3}$\thanks{E-mail: borko@electra.bajaobs.hu},
S. A. Rappaport$^4$,
P.\,F.\,L.\,Maxted$^5$
I.~Terentev$^{6}$,
M.\,Omohundro$^{6}$,
\newauthor
R.~Gagliano$^7$,
T.\,Jacobs$^8$,
M.\,H.\,Kristiansen$^{9,10}$,
D.\,LaCourse$^{11}$,
H.\,M.\,Schwengeler$^6$,
\newauthor
A.\,Vanderburg$^{12}$,
M.\,G.\,Blackford$^{13}$
\\
$^{1}$Baja Astronomical Observatory of University of Szeged, H-6500 Baja, Szegedi \'ut, Kt. 766, Hungary\\
$^{2}$Konkoly Observatory, Research Centre for Astronomy and Earth Sciences,  H-1121 Budapest, Konkoly Thege Mikl\'os \'ut 15-17, Hungary\\
$^3$ ELTE Gothard Astrophysical Observatory, H-9700 Szombathely, Szent Imre h. u. 112, Hungary \\
$^4$ Department of Physics, Kavli Institute for Astrophysics and Space Research, M.I.T., Cambridge, MA 02139, USA\\
$^5$ Astrophysics Group, Keele University, Staffordshire, ST5 5BG, UK\\
$^6$ Citizen Scientist, c/o Zooniverse, Department of Physics, University of Oxford, Denys Wilkinson Building, Keble Road, Oxford, OX1 3RH, UK \\
$^7$ Amateur Astronomer, Glendale, AZ 85308 \\
$^8$ Amateur Astronomer, 12812 SE 69th Place Bellevue, WA 98006, USA \\
$^{9}$ Brorfelde Observatory, Observator Gyldenkernes Vej 7, DK-4340 T\o ll\o se, Denmark \\
$^{10}$ DTU Space, National Space Institute, Technical University of Denmark, Elektrovej 327, DK-2800 Lyngby, Denmark \\
$^{11}$ Amateur Astronomer, 7507 52nd Place NE Marysville, WA 98270, USA \\
$^{12}$ Department of Astronomy, The University of Wisconsin-Madison, 475 N.\,Charter St., Madison, WI 53706, USA \\
$^{13}$ Congarinni Observatory, Congarinni, NSW, Australia \\
}

\date{Accepted XXX. Received YYY; in original form ZZZ}

\pubyear{2021}

\begin{document}
\label{firstpage}
\pagerange{\pageref{firstpage}--\pageref{lastpage}}
\maketitle

\begin{abstract}
BG\,Ind is a well studied, bright, nearby binary consisting of a pair of F stars in a 1.46-day orbit.  We have discovered in the \textit{TESS} lightcurve for TIC 229804573 (aka BG\,Ind) a second eclipsing binary in the system with a $0.53$-day.  Our subsequent analyses of the recent {\em TESS} and archival ground-based photometric and radial velocity data, reveal that the two binaries are gravitationally bound in a 721-day period, moderately eccentric orbit.  We present the results of a joint spectro-photodynamical analysis of the eclipse timing variation curves of both binaries based on {\em TESS} and ground-based archival data, the \textit{TESS} lightcurve, archival radial velocity data and the spectral energy distribution, coupled with the use of PARSEC stellar isochrones. We confirm prior studies of BG\,Ind which found that the brighter binary A consists of slightly evolved F-type stars with refined masses of 1.32 and 1.43 $M_\odot$, and radii of 1.59 and 2.34 $R_\odot$. The previously unknown binary B has two less massive stars of 0.69 and 0.64 $M_\odot$ and radii of 0.64  and 0.61 $R_\odot$. Based on a number of different arguments which we discuss, we conclude that the three orbital planes are likely aligned to within 17$^\circ$. 
\end{abstract}




\section{Introduction}


BG~Ind ($\kappa_1$~Ind; HD\,208496; TIC\,229804573) is a bright, sixth magnitude eclipsing binary (EB) formed by two F-type stars. Its variability was reported first by \citet{strohmeieretal64}, and its EB nature was found by \citet{manfroidmathys84}. At the same time, \citet{andersenetal84} obtained the first two spectrograms, and concluded that BG~Ind is also a double-lined spectroscopic binary and calculated stellar masses and radii for the first time.  The first photometric lightcurve analysis was carried out by \citet{vanhammemanfroid88}. 

In the forthcoming decades several new photometric and spectroscopic observations were carried out. They are nicely summarized in \citet{rozyczkaetal11}, and therefore we do not repeat them here. 

The most recent thorough spectroscopic and photometric analysis was carried out by \citet{rozyczkaetal11}. These authors analysed all the available lightcurves and radial velocity (RV) data including their own measurements. They performed extensive spectroscopic analyses to obtain accurate stellar temperatures, system abundances and then age and evolutionary status. We will compare their results with our findings later in Sect.~\ref{sec:discussion}, and therefore, here we highlight only a few noteworthy details. First, they found that the more massive and larger star has the lower temperature\footnote{In most binaries with unevolved stars the primary star is the more massive and the hotter star.  In the A binary of this system (the dominant binary), the more massive star turns out to be the cooler of the two due to its evolution.  We continue to refer to the more massive star as the `primary' even though it is cooler.  However, we will still refer to the deeper eclipse (i.e., when the primary eclipses the less massive (but hotter star) as the `primary eclipse'.  The usual naming convention still holds for the fainter binary B.}, thereby indicating clearly that this component has already evolved away from the main sequence and is moving toward the subgiant regime. Second, they made attempts to resolve some problems with both the photometric phasing (already first noted in \citealp{vanhammemanfroid88}) and discrepancies in the systemic $\gamma$ velocities obtained in the solutions of the RV curves measured during three highly different epochs by \citet{andersenetal84,bakisetal10,rozyczkaetal11}.  However, they were not able arrive at any definitive conclusions regarding these inconsistencies. 



We further note that BG\,Ind was included in the catalog of those detached eclipsing binaries for which the constituent masses and radii are known to at least 2\% precision \citep{southworth15}.  And BG\,Ind was also selected for inclusion in the sample of 156 detached eclipsing binaries, which can be used as benchmarks for trigonometric parallaxes in the Gaia era \citep{stassuntorres16}. Finally, turning to the Gaia era, with the use of Gaia DR2 \citep{GaiaDR2} and HIPPARCOS \citep{HIPrev} data, a significant proper motion anomaly was found that might indicate the presence of further, gravitationally bound components in the system \citep{brandt18,kervellaetal19}.  At this point it should also be noted, that there is a remarkable discrepancy between the revised HIPPARCOS and Gaia EDR3 \citep{GaiaEDR3} parallaxes of BG~Ind ($\pi_\mathrm{HIP}=14.90\pm0.59$\,mas vs $\pi_\mathrm{EDR3}=19.44\pm0.52$\,mas), which might be a further indicator of additional multiplicity in the system. 

In this paper we confirm the---at least---quadruple nature of BG Ind. Using the high-precision {\em TESS} photometry with 2-min cadence we have discovered an obvious second eclipsing EB in the lightcurve of BG\,Ind with a period of 0.53 days\footnote{This second eclipsing binary was also found independently by \citealp{eisneretal21}.}. Our comprehensive investigation of the \textit{TESS} photometry, archival ground-based photometry and radial velocity curves, as well as the eclipse timing variations (ETV) data demonstrates that the two eclipsing binaries form a close 2+2 quadruple stellar system with a remarkably short outer period of $\sim2$ years.

In Sect.~\ref{sec:obsdata} we describe all the available observational data and their preparation for the complex, joint photodynamical analysis which is discussed in Sect.~\ref{Sect:photodyn}. Then, the results are discussed and finally, summarized in Sects.~\ref{sec:discussion} and \ref{Sect:Summary}.

\section{Observational data}
\label{sec:obsdata}

\subsection{Catalog data}
\label{subsect:catalogdata}

In Table~\ref{tab:catalogs}, in addition to other catalog data, we collected the photometric passband magnitudes of the system from different surveys, e.g. Tycho-2 \citep{Tycho2}, 2MASS \citep{2MASS}, AllWISE \citep{WISE}, GALEX \citep{GALEX} and Gaia \citep{GaiaEDR3}.  These will be used to construct the spectral energy distribution (`SED') of the system. In turn, the SED along with theoretical isochrones and the photodynamical model of the system provide an opportunity to determine the masses of the components in an astrophysical model-dependent way (see Sect.~\ref{Sect:photodyn} for details). Together with the passband magnitudes given in Table~\ref{tab:catalogs}, we list their uncertainties as tabulated in the given catalogs. For the SED analysis, however, we used a minimum uncertainty of $0.03$ mag to avoid the overdominance of the extremely precise Gaia magnitudes and also to counterbalance the  uncertainties inherent in our interpolataion method during the calculations of theoretical passband magnitudes that are part of the fitting process. Furthermore, similar to the approach followed by \citet{stassuntorres16} we omitted the GALEX near-UV magnitude from our analysis as a distinct outlier. K. Stassun (private communication) kindly called our attention to the fact that even the largest available NUV aperture is missing flux.

\begin{table}
	\centering
	\caption{Main properties of BG\,Ind from different catalogs.}
	\label{tab:catalogs}
	\begin{tabular}{c c c}
		\hline
		Parameter & Value & References\\
		\hline
		RA & 329.62537 & 1\\
		DEC & -59.01201 & 1\\
		$\mu_{\mathrm{RA}}$ [mas\,yr$^{-1}$] & 4.96 $\pm$ 0.35  & 1\\
		$\mu_{\mathrm{DEC}}$ [mas\,yr$^{-1}$] & 30.21 $\pm$ 0.53 & 1\\
                $\pi_\mathrm{EDR3}$ [mas] & $19.44\pm0.52$ & 1 \\ 
                $\pi_\mathrm{HIP}$ [mas] & $14.90\pm0.59$ & 2 \\
		G & 6.024606 $\pm$ 0.001610 & 1\\
		G$_{BP}$ & 6.266601 $\pm$ 0.005111 & 1\\
		G$_{RP}$ & 5.645544 $\pm$ 0.010610 & 1\\
		T & 5.6502 $\pm$ 0.0067 & 3\\
		B & 6.605 $\pm$ 0.022 & 3\\
		V & 6.130 $\pm$ 0.030 & 3\\
                B$_T$& 6.697 $\pm$ 0.014 & 4\\
                V$_T$& 6.195 $\pm$ 0.009 & 4\\
		J & 5.206 $\pm$ 0.020 & 5\\
		H & 4.993 $\pm$ 0.026 & 5\\
		K & 4.877 $\pm$ 0.026 & 5\\
		W1 & 4.907 $\pm$ 0.215 & 6\\
		W2 & 4.615 $\pm$ 0.092 & 6\\
                W3 & 4.897 $\pm$ 0.014 & 6\\
                W4 & 4.821 $\pm$ 0.028 & 6\\
                FUV&15.216 $\pm$ 0.015 & 7\\
                NUV&11.598 $\pm$ 0.002 & 7\\
                $[M/H]$ [dex] & $-0.30$ & 8\\
		Distance [pc] &  $51.0\pm0.5$  & 9\\
		\hline
	\end{tabular}
	
\textbf{References. }(1) Gaia EDR3 \citep{GaiaEDR3}; (2) HIPPARCOS (revised) \citep{HIPrev}; (3) TIC-8 catalog \citep{TIC}; (4) Tycho-2 catalog \citep{Tycho2}; (5) 2MASS All-Sky Catalog of Point Sources \citep{2MASS}; (6) AllWISE catalog \citep{WISE}; (7) GALEX-DR5 (GR5) \citep{GALEX}  ;(8) \citet{holmbergetal09}; (9) \citet{bailer-jonesetal18}
\end{table}


\subsection{\textit{TESS} photometry}
\label{sec:TESSphot} 

\begin{figure}
\begin{center}
\includegraphics[width=0.47 \textwidth]{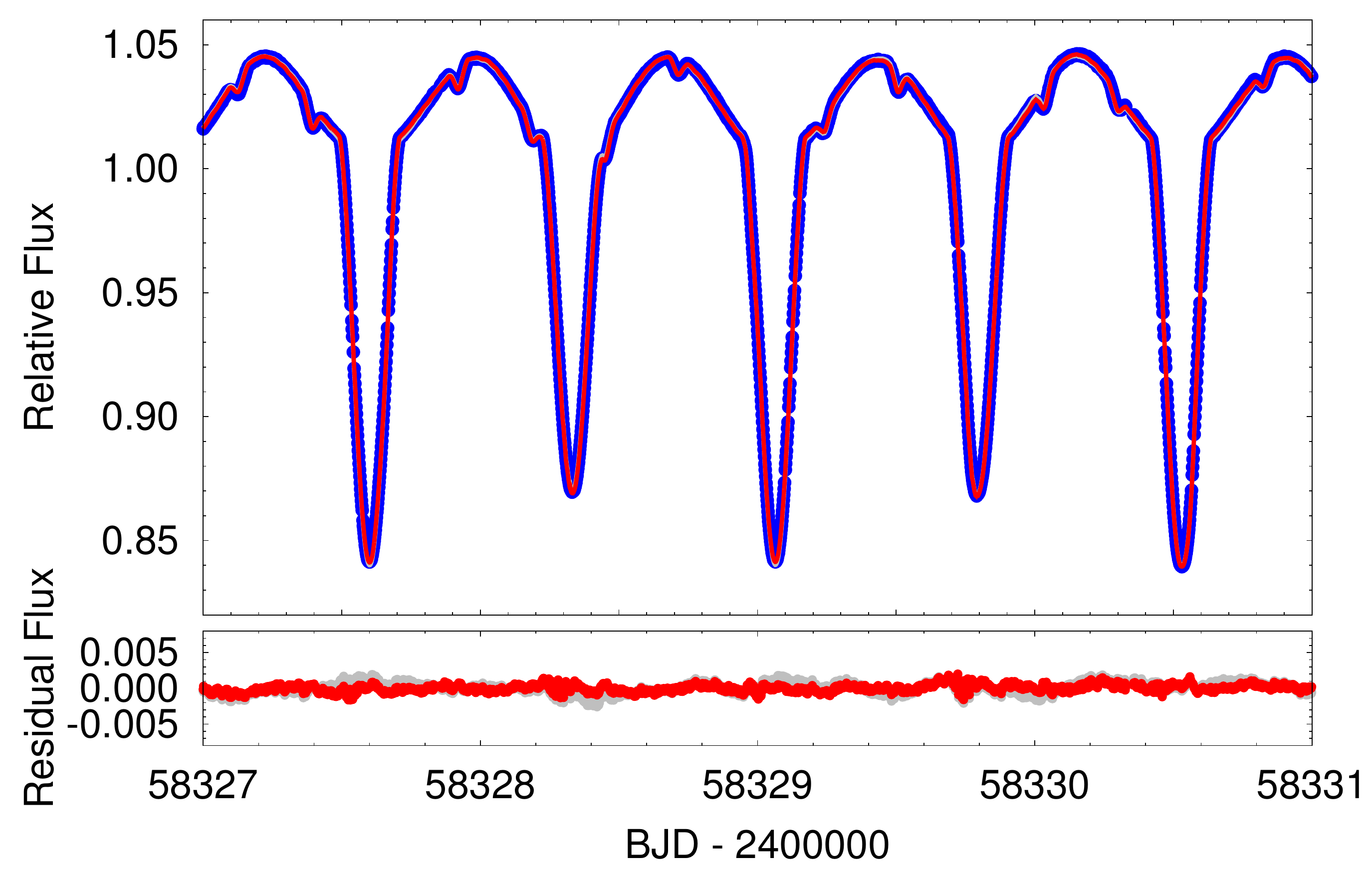}
\includegraphics[width=0.47 \textwidth]{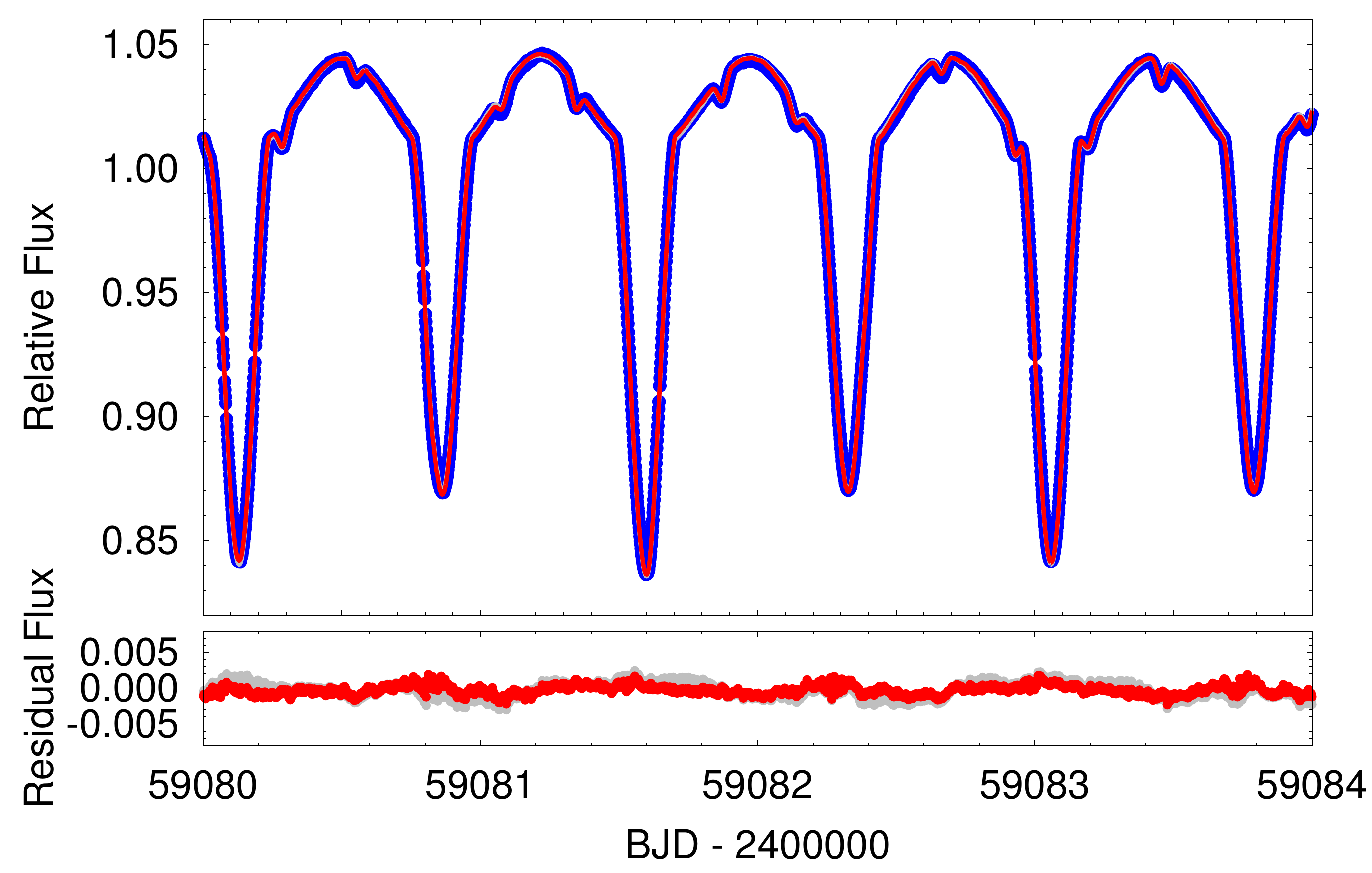}
\caption{Two four-day sections from the beginning of Sector 1 and the end of Sector 28 SAP lightcurves of BG\,Ind (blue circles). The red and grey curves are spectro-photodynamical model solutions (see later, in Sect.~\ref{Sect:photodyn}). In the case of the red solution the small extra fluctuations of the lightcurve are probably due to the chromospheric/photospheric activities of the stars and were modelled mathematically with Fourier-harmonics simultaneously with the two-binary model, while the grey curve represent the pure two-binary part of the same solution. The residuals to the models are also shown below the lightcurves.} 
\label{fig:lightcurve} 
\end{center}
\end{figure}

The \textit{TESS} space telescope \citep{ricker15} has observed this target in 2-min cadence mode during Sectors 1, 27 and 28. We downloaded both the Simple Aperture Photometry (SAP) and the Pre-search Data Conditioning SAP (PDCSAP) lightcurves from the MAST portal\footnote{\href{https://mast.stsci.edu}{https://mast.stsci.edu}}. We used the SAP lightcurves for our study. Because the presence of the small extra dips belonging to the eclipses of the previously unknown binary B (see Fig.\,\ref{fig:lightcurve}) was discovered shortly after the release of the data of the first four {\em TESS} sectors, our analyses were carried out mostly with the use of Sector 1 data. We also did use Sector 27 and 28 data, but mainly for the purpose of extending the interval of the ETV study.  Since, in the case of the faint binary B, the only sources of ETV data are the three sectors of high-quality \textit{TESS} data, the inclusion of these new observations into our analysis significantly improved the accuracy of the outer orbit solution (including the dynamically inferred mass of binary B).

\subsection{WASP photometry}

BG~Ind is one of millions of stars that have been observed as part of the WASP survey. The survey is described in \citet{2006PASP..118.1407P} and \citet{2006MNRAS.373..799C}. From July 2012  the  WASP-South  instrument  was operated using  85-mm, f/1.2 lenses and an r$^{\prime}$ filter. With these lenses the image scale is 33 arcsec/pixel. Observations of BG~Ind were obtained simultaneously in two cameras on WASP-South over three observing seasons, from 2012 July 3  to 2014 December 6. Fluxes are measured in an aperture with a radius  of 132 arcsec for the 85-mm data and instrumental trends are removed using the SYSRem algorithm \citep{2005MNRAS.356.1466T}. Data points more than 5 standard deviations from a phase-binned version of the light curve were rejected and the entire night of data was rejected if more than 1/4 of the observations were identified as outliers based on this criterion. 

\begin{figure}
\begin{center}
\includegraphics[width=0.47 \textwidth]{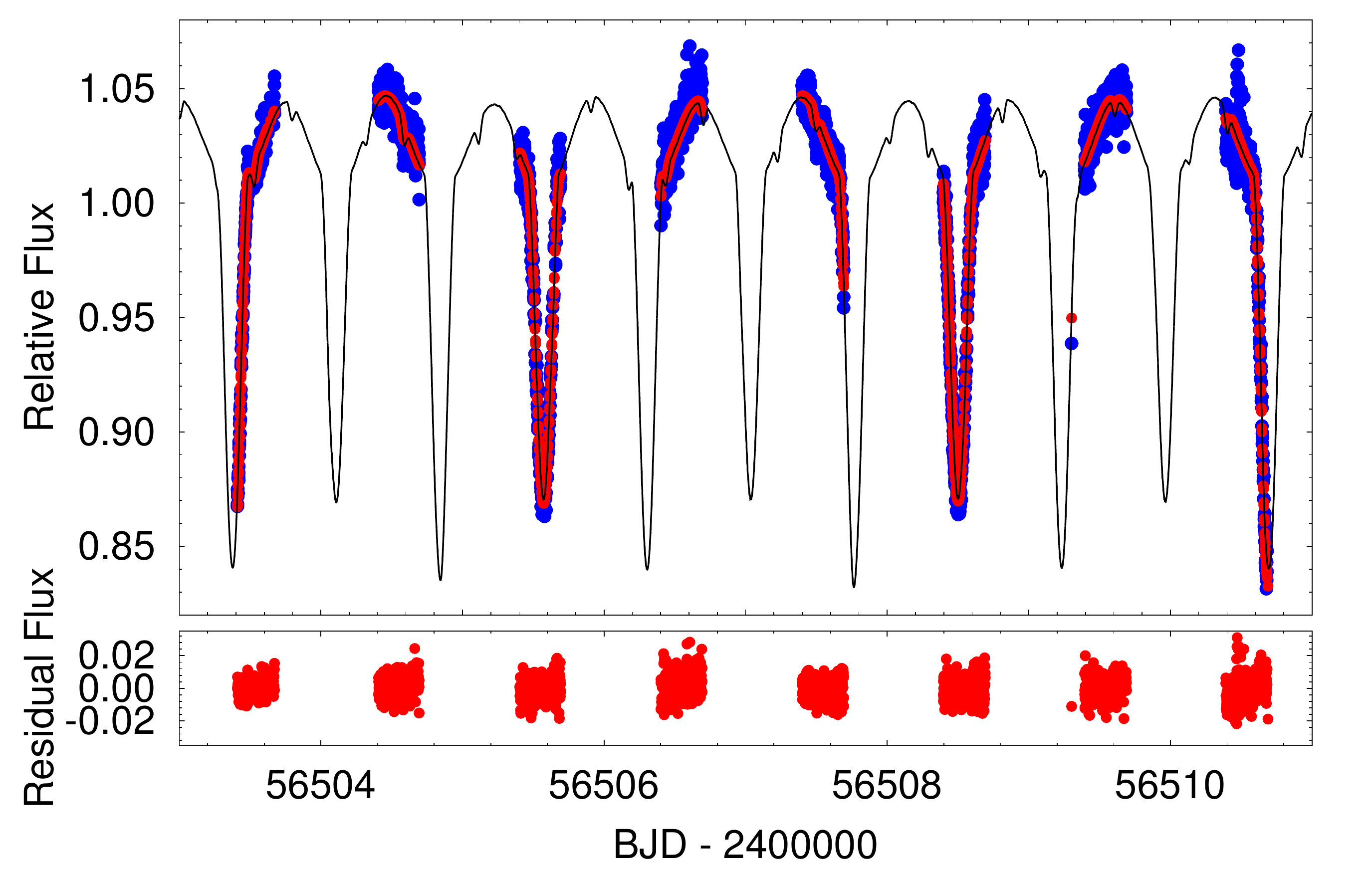}
\caption{An eight-day-long section of the WASP observations of BG\,Ind (blue circles). Red dots represent the best-fit spectro-photodynamical model solution (see later, in Sect.~\ref{Sect:photodyn}) projected back to the epoch of each individual WASP measurement, while the black line shows the evenly phased model solution. The residuals curve is also shown below.} 
\label{fig:wasp_lc} 
\end{center}
\end{figure}

An eight-day section of the WASP measurements is shown in Fig.~\ref{fig:wasp_lc}. Note, we converted the original HJD(UTC) times of the WASP observations to BJD(TDT) for the forthcoming analyses. We also applied the same transformations for all the archival data that we describe below.

\subsection{Other ground-based archive photometric data used for our analysis}
\label{subsect:lc_ESO_archive}
We downloaded publicly available Str\"omgren $u$, $v$, $b$, $y$ photometric observations from the ESO archive \citep{manfroidetal91,sterkenetal93}. These observations were carried out between JDs 2\,446\,581 and 2\,447\,069. The lightcurves in each bandpass contain 175 measurements. We used these data primarily to determine additional times of eclipses in binary A. Unfortunately, however, the majority of the nightly observations contain only a few measurements, and we were therefore able to determine the mid-times of only two primary eclipses (see below, in Sect.\,\ref{subsect:ETVdata}) from this data set. 

BG Ind was also observed by Jens Viggo Clausen and collaborators as part of their long-running observing programme to measure absolute dimensions for solar-type stars in eclipsing binaries, carried out since 1994 at the Str{\"o}mgren Automatic Telescope at ESO, La Silla. Unpublished data and manuscripts that were in preparation from this observing programme have been made available to one of us (PM), from which we extracted 5 more eclipse times (see again, in Sect.\,\ref{subsect:ETVdata}). The observing procedures and data reduction for these observations are similar to those described in \citet{clausenetal01}.

\subsection{Disentanglement of the lightcurves}
\label{subsect:disentanglement}

\begin{figure}~
\begin{center}
\includegraphics[width=0.47 \textwidth]{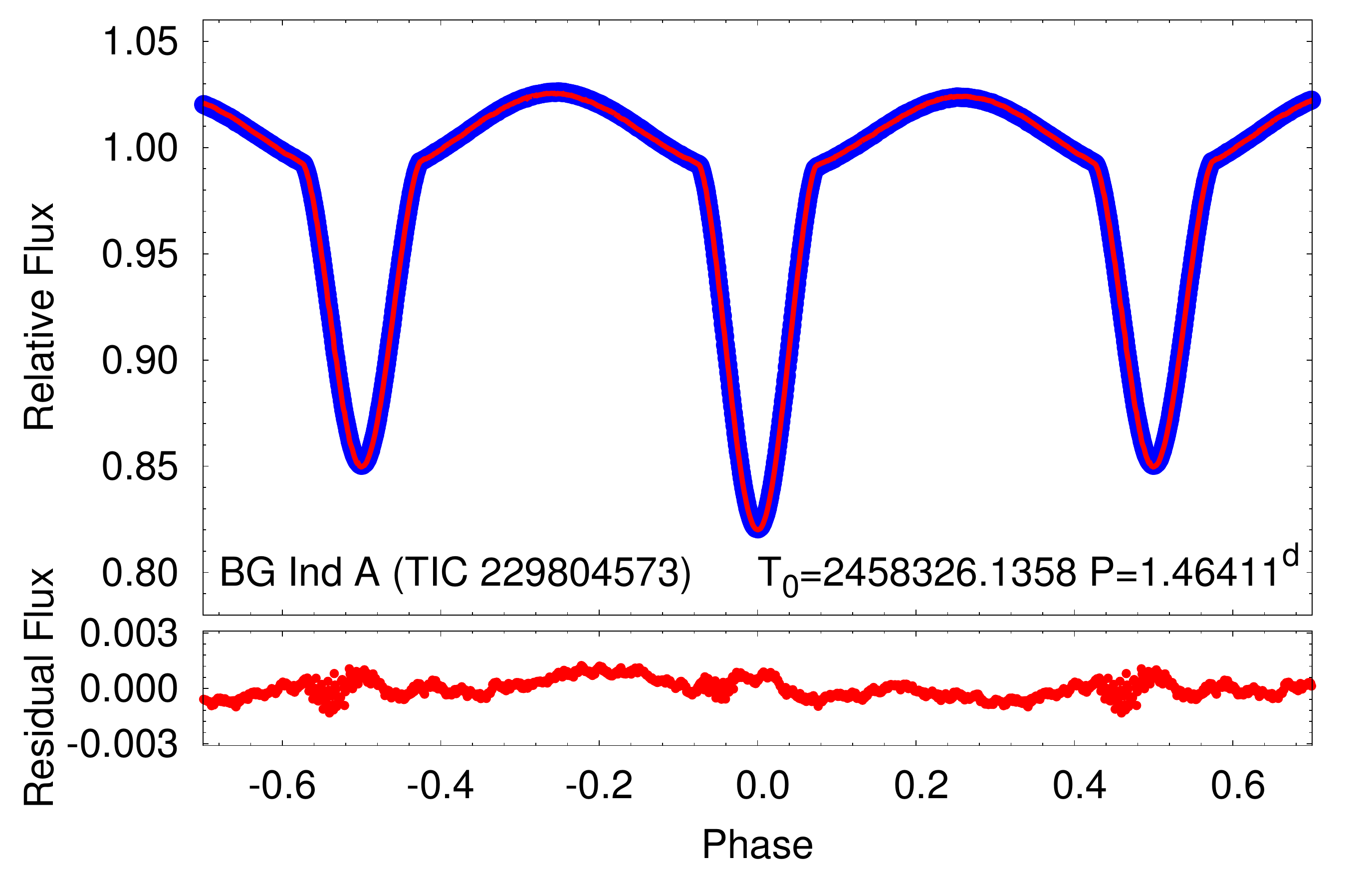}
\includegraphics[width=0.47 \textwidth]{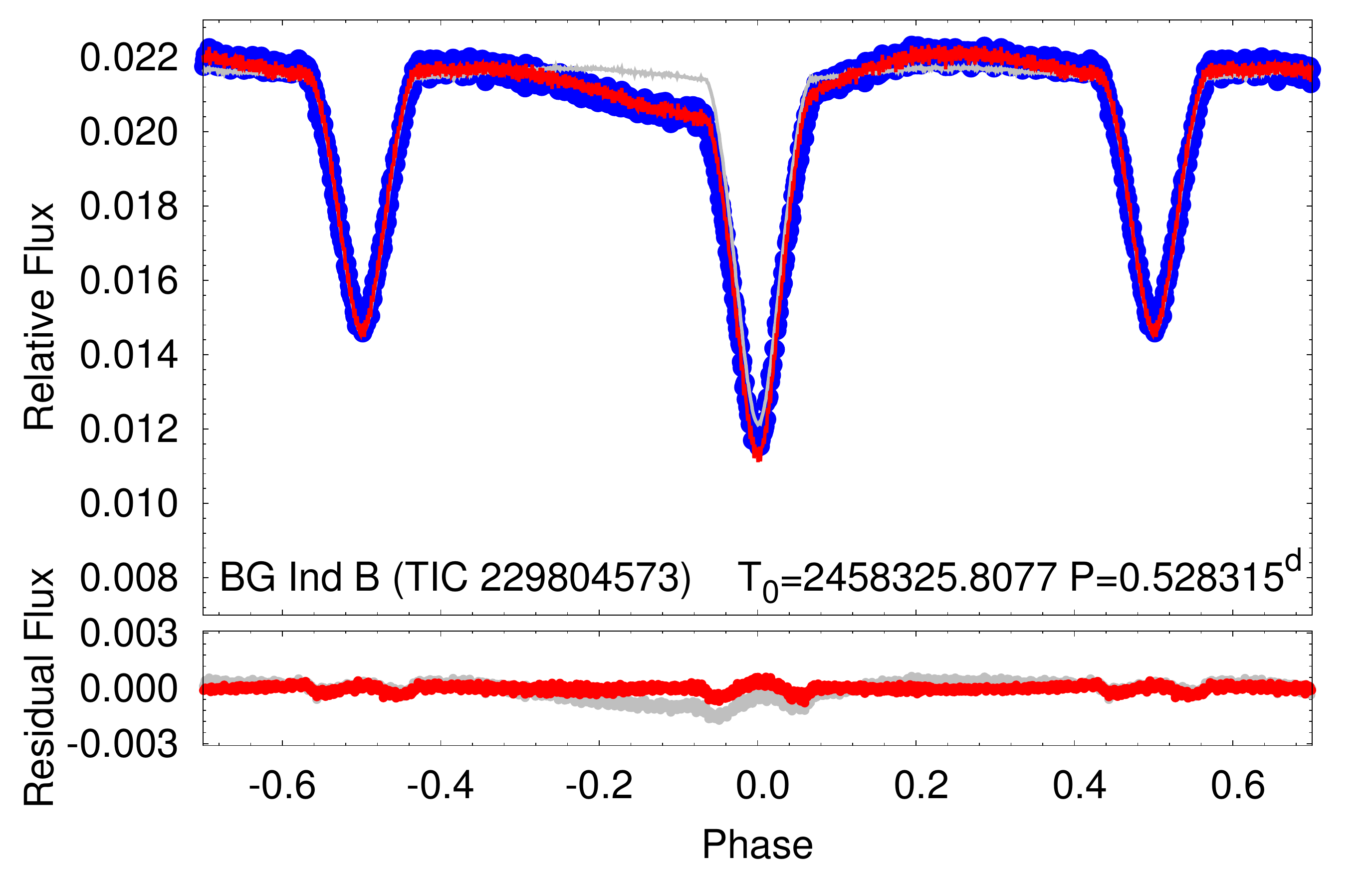}
\includegraphics[width=0.47 \textwidth]{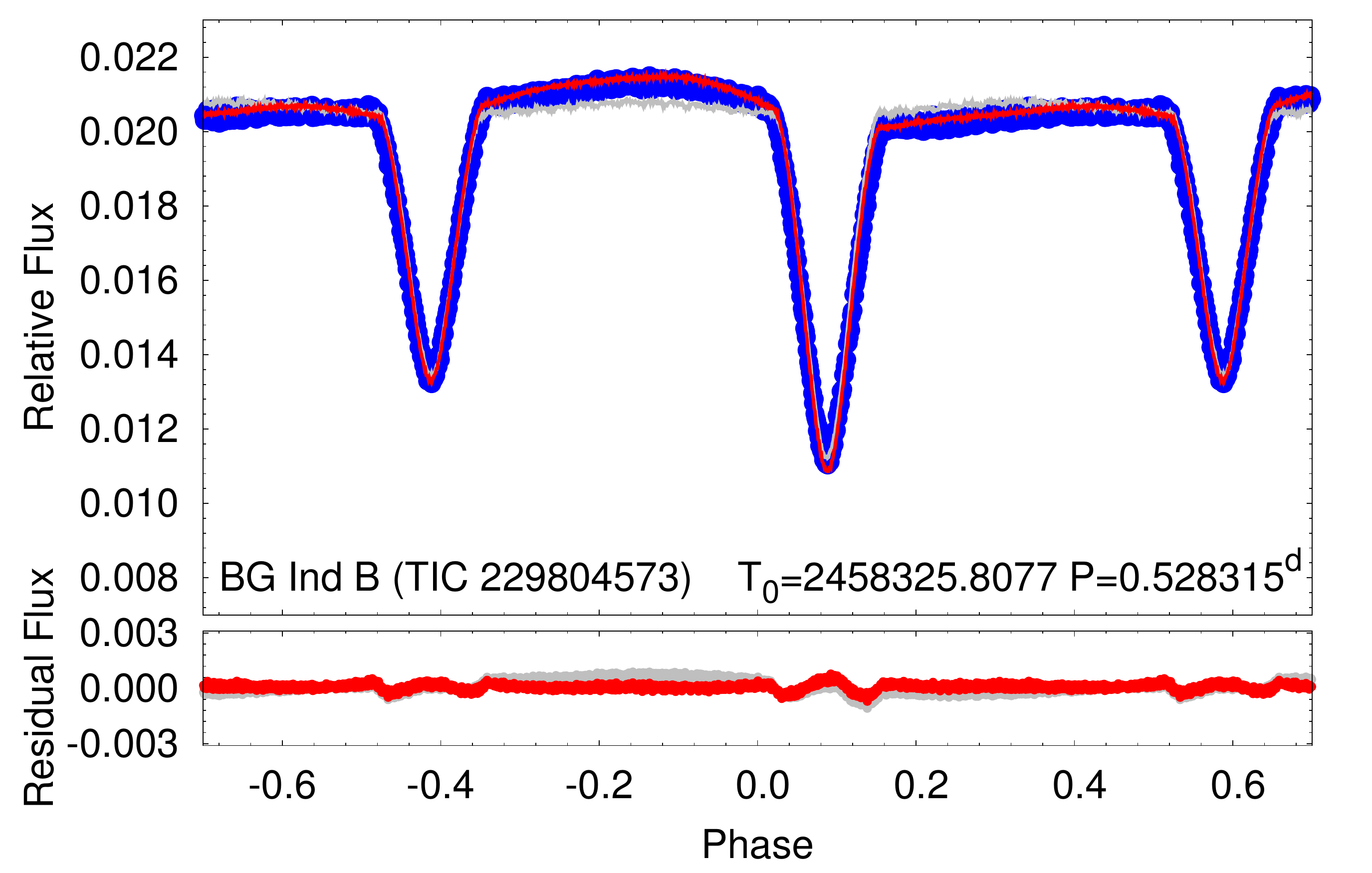}
\caption{Folded, binned, and averaged \textit{TESS} lightcurves of the two binaries of the quadruple system BG~Ind. {\it Upper panel:} Sector 1 lightcurve of binary A (blue circles), together with the folded, binned, and averaged combined spectro-photodynamical model lightcurve (red curve, see later, in Sect.~\ref{Sect:photodyn}). {\it Middle and lower panels:} Year 1 (Sector 1) and Year 3 (Sector 27 and 28) lightcurves of binary B, respectively. As in the case of the binary A lightcurve in Fig.~\ref{fig:lightcurve}, the red solution curve exhibits some small extra fluctuations that are probably due to the chromospheric/photospheric activities of the stars (see Sect.~\ref{Sect:photodyn} for details). These were modelled mathematically with Fourier-harmonics simultaneously with the two-binary model, while the thin grey curves represent the pure two-binary part of the same solution. The fold of the residuals to the models are also shown below the folded lightcurves.}
\label{fig:ABfold} 
\end{center}
\end{figure}

For the combined analysis of all the observational data we used the original \textit{TESS} timeseries, i.e., the net lightcurve of the two binaries together. However, at the start of the analysis, we found it worthwhile to disentangle the lightcurves of the two binaries, so we could examine each one separately.  We have described this process in substantial detail in \citet{powelletal21}.  Therefore we review only the highlights here.  First we folded and binned the Sector 1 (i.e. Year 1)  \textit{TESS} SAP lightcurve with the period of binary A into 1000 equal phase cells.  However, while producing the fold for binary A, we excluded those data points which were recorded during the eclipses of binary B.  Then the mean flux of each cell was rendered to the mid-phase value of that cell.  In such a way we obtained a folded, binned, and averaged lightcurve of binary A (see upper panel of Fig.~\ref{fig:ABfold}). Then this lightcurve was removed from the original sector 1 SAP lightcurve point by point in such a manner that the flux to be removed at the actual phase of any given data point was calculated with a three-point local Lagrange-interpolation from the folded, binned, and averaged lightcurve. As the result of this removal, we have obtained a new, residual time series which now mainly contains the light variations of binary B,\footnote{Note, for practical reasons, we added a constant flux to these time series in such a way that the flux of the very first data point retained the same value as in the original time series. In this manner, we replaced the varying light of the extracted binary with a constant extra light.} without the eclipses and ellipsoidal variations of binary A. Therefore, this lightcurve can be used for determining the mid-eclipse times of binary B.

In the next step we folded, binned, and averaged this residual lightcurve with the period of binary B (see middle panel of Fig.~\ref{fig:ABfold}). Finally, we subtracted this folded, disentangled lightcurve of binary B from the original Sector 1 \textit{TESS} SAP lightcurve, thereby obtaining a time series of binary A without the small distortions caused by binary B. 
We applied the same process to the Sector 27 and 28 (Year 3) SAP lightcurves, as well (see bottom panel of Fig.~\ref{fig:ABfold}).

\begin{figure}
\begin{center}
\includegraphics[width=0.47 \textwidth]{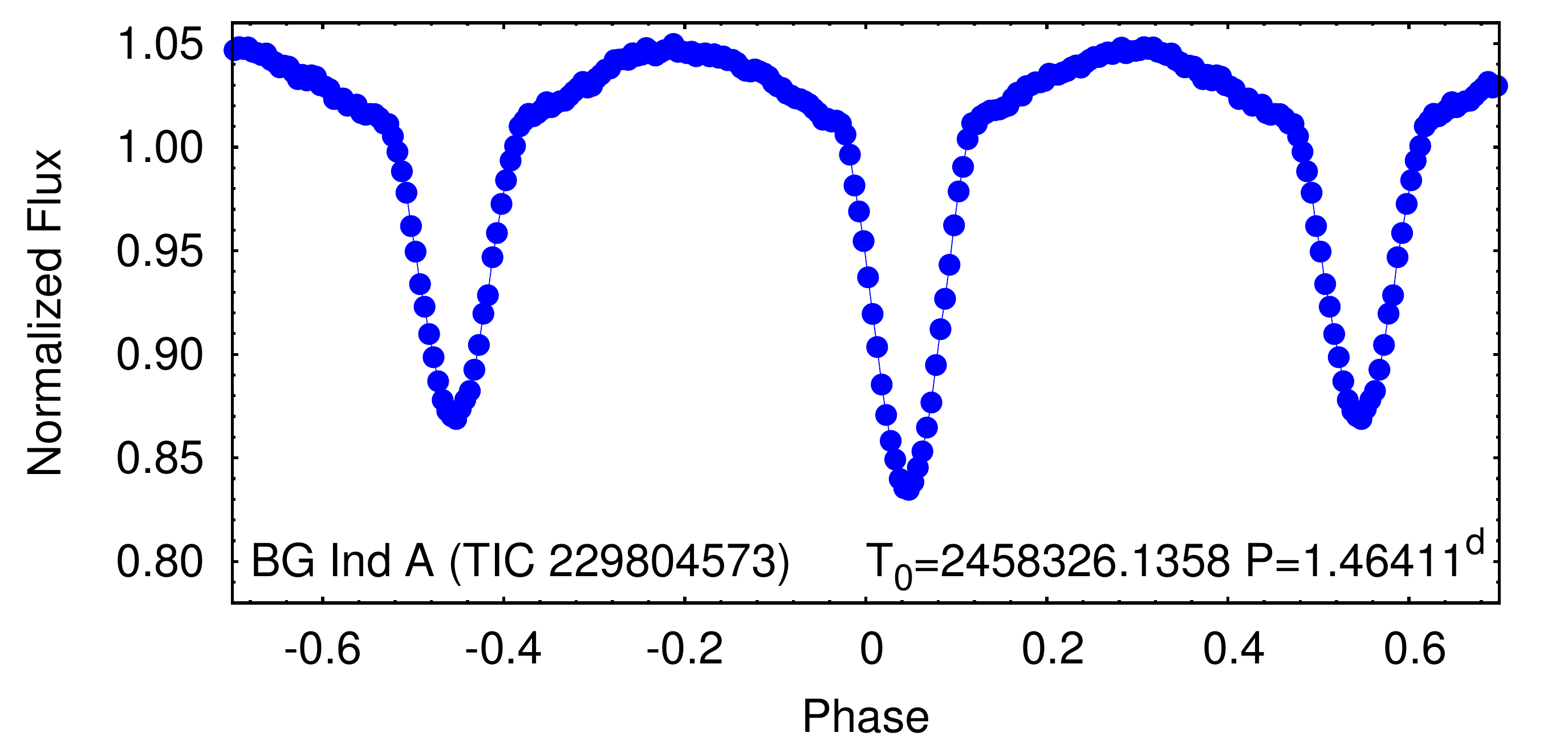}
\includegraphics[width=0.47 \textwidth]{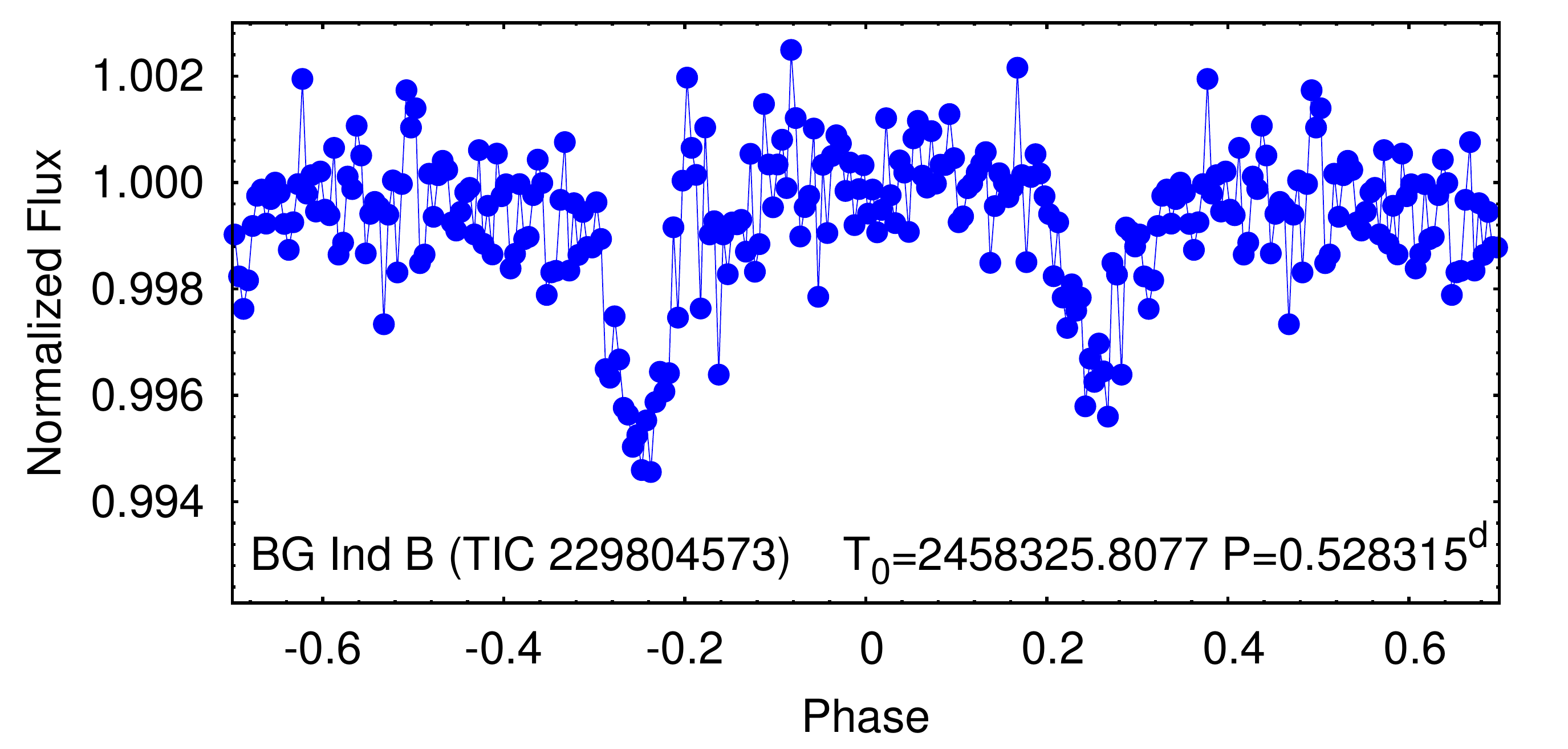}
\caption{Folded, binned, averaged WASP lightcurves of binary A and B for observing season 2012/2013. For illustrative purposes we phased both curves with the ephemeris calculated for Sector 1 \textit{TESS} data. In such a manner, the shift of the primary and secondary eclipses from phases $0\fp0$ and $0\fp5$, respectively, i.e., the phasing problem mentioned in the Introduction, is clearly visible.} 
\label{fig:WASPfold} 
\end{center}
\end{figure}

Regarding the WASP observations, we carried out a very similar process with the slight modification that, in this case, for the much smaller number of individual data points we applied binnings of 200 and 500 cells instead of 1000. We carried out the whole process separately for the three seasons of the WASP observations. Though the eclipsing signal of the faint binary B is not readily detected in the original WASP timeseries, we were able to see it clearly in our disentangled version (see Fig.~\ref{fig:WASPfold}).

\begin{figure}
\begin{center}
\includegraphics[width=0.47 \textwidth]{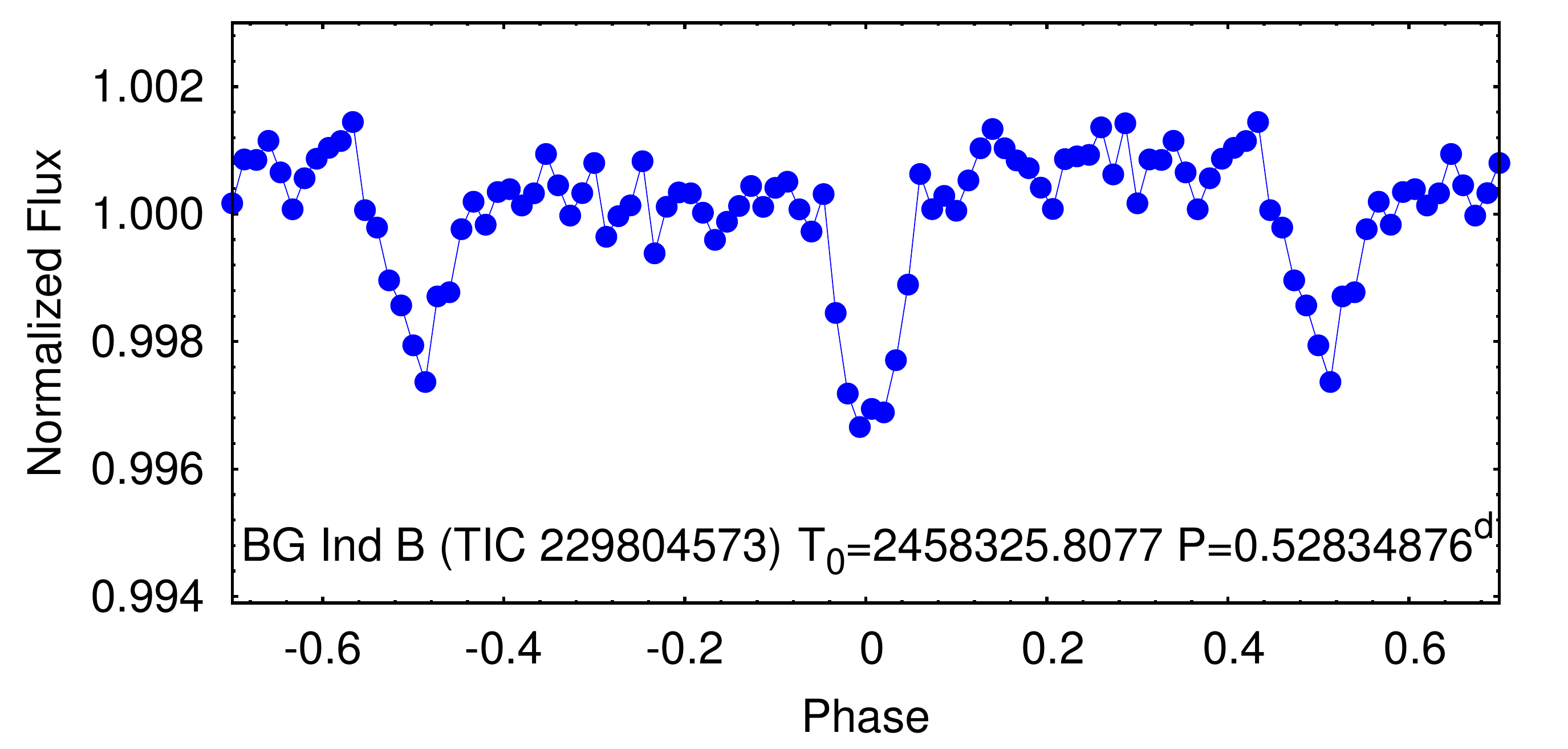}
\caption{The folded, binned, averaged WASP lightcurve of binary B obtained with the use of the second disentanglement method based on Fourier-filtering. The full, three-season WASP lightcurve was folded with the mean orbital period obtained from the ETV analysis (see Sect.~\ref{Sect:photodyn}).} 
\label{fig:compare} 
\end{center}
\end{figure}

Finally in regard to disentangling the lightcurves, we have also used a second method which fits simultaneously for 50 harmonics of each of binaries A and B given their established periods.  This technique, which is also described in detail in \citet{powelletal21}, involves inverting a $201 \times 201$ matrix to solve for the linear coefficients to the 50 sines and 50 cosines for each of the two binaries.  We find nearly perfect agreement for the disentangled {\em TESS} lightcurves from the two independent methods, and thus we do not show those results here.  In the case of disentangling the WASP data, the results for binary B are actually somewhat improved using the Fourier approach and we show that lightcurve in Fig.~\ref{fig:compare} as well for comparison.

\subsection{ETV data}
\label{subsect:ETVdata}

\subsubsection{{\em TESS} ETV results}

In order to calculate accurate eclipse times from the \textit{TESS} lightcurve we used the disentangled time series (see above in Sect.\,\ref{subsect:disentanglement}).  The 91 eclipse times of binary A (including both primary and secondary eclipses) from Sectors 1, 27, and 28 are presented in Table \ref{tab:BGIndA_etv_TESS}.  In Table \ref{tab:BGIndB_etv_TESS} we list the eclipse times for binary B including a combined 259 primary and secondary eclipses.

\subsubsection{Ground-based ETV results}

We also utilized WASP and ESO \citep{manfroidetal91,sterkenetal93} data, including the unpublished observations of J.\,V.\,Clausen, as well, to calculate 85 additional eclipse times for binary A.  Furthermore, a primary and a secondary eclipse of BG~Ind were observed by one of us (MB) using a DSLR camera. Images were recorded in RAW format and the green, blue and red channels were extracted into separate images. The times of minimum were measured from each colour filter with the {\sc Peranso} software\footnote{\url{https://www.cbabelgium.com/peranso/}} using a 5th order polynomial fit. The average of the mid-eclipse times were converted into BJD. Finally, we collected one other eclipse time from the the paper of \citet{vanhammemanfroid88} and converted it into BJD.  All these eclipse times are tabulated in Table~\ref{tab:BGIndA_etv_ground}. The eclipses from binary B in the archival data were too weak to derive meaningful eclipse times.

\begin{table*}
\caption{Eclipse times of BG Ind binary A determined from \textit{TESS} observations.}
 \label{tab:BGIndA_etv_TESS}
\begin{tabular}{@{}lrllrllrl}
\hline
BJD & Cycle  & std. dev. & BJD & Cycle  & std. dev. & BJD & Cycle  & std. dev. \\ 
$-2\,400\,000$ & no. &   \multicolumn{1}{c}{$(d)$} & $-2\,400\,000$ & no. &   \multicolumn{1}{c}{$(d)$} & $-2\,400\,000$ & no. &   \multicolumn{1}{c}{$(d)$} \\ 
\hline
58325.403840 & -0.5   &  0.000005  &  58349.561705 & 16.0   &  0.000005  &  59056.704428 & 499.0  &  0.000005 \\
58326.135712 & 0.0    &  0.000004  &  58350.293558 & 16.5   &  0.000004  &  59057.436384 & 499.5  &  0.000005 \\
58326.867912 & 0.5    &  0.000005  &  58351.025742 & 17.0   &  0.000006  &  59058.168508 & 500.0  &  0.000005 \\
58327.599837 & 1.0    &  0.000003  &  58351.757721 & 17.5   &  0.000007  &  59058.900490 & 500.5  &  0.000006 \\
58328.332109 & 1.5    &  0.000005  &  58352.489843 & 18.0   &  0.000004  &  59059.632635 & 501.0  &  0.000005 \\
58329.063938 & 2.0    &  0.000004  &  59036.939185 & 485.5  &  0.000005  &  59060.364549 & 501.5  &  0.000007 \\
58329.796203 & 2.5    &  0.000005  &  59037.670779 & 486.0  &  0.000005  &  59062.560877 & 503.0  &  0.000004 \\
58330.528118 & 3.0    &  0.000004  &  59038.403292 & 486.5  &  0.000006  &  59063.292963 & 503.5  &  0.000005 \\
58331.260286 & 3.5    &  0.000006  &  59039.134927 & 487.0  &  0.000004  &  59064.024900 & 504.0  &  0.000004 \\
58331.992218 & 4.0    &  0.000004  &  59039.867263 & 487.5  &  0.000006  &  59064.757088 & 504.5  &  0.000004 \\
58332.724303 & 4.5    &  0.000006  &  59040.598986 & 488.0  &  0.000005  &  59065.489098 & 505.0  &  0.000004 \\
58333.456408 & 5.0    &  0.000004  &  59041.331371 & 488.5  &  0.000006  &  59066.221307 & 505.5  &  0.000005 \\
58334.188441 & 5.5    &  0.000005  &  59042.063196 & 489.0  &  0.000005  &  59066.953158 & 506.0  &  0.000004 \\
58334.920563 & 6.0    &  0.000005  &  59042.795556 & 489.5  &  0.000006  &  59067.685476 & 506.5  &  0.000006 \\
58335.652617 & 6.5    &  0.000005  &  59043.527276 & 490.0  &  0.000005  &  59068.417320 & 507.0  &  0.000003 \\
58336.384671 & 7.0    &  0.000004  &  59044.259642 & 490.5  &  0.000006  &  59069.149723 & 507.5  &  0.000007 \\
58337.116773 & 7.5    &  0.000005  &  59044.991370 & 491.0  &  0.000005  &  59069.881452 & 508.0  &  0.000004 \\
58337.848766 & 8.0    &  0.000005  &  59045.723835 & 491.5  &  0.000006  &  59070.613800 & 508.5  &  0.000007 \\
58340.044827 & 9.5    &  0.000005  &  59046.455589 & 492.0  &  0.000005  &  59071.345545 & 509.0  &  0.000004 \\
58340.777035 & 10.0   &  0.000004  &  59047.187905 & 492.5  &  0.000006  &  59075.737851 & 512.0  &  0.000004 \\
58341.508927 & 10.5   &  0.000006  &  59047.919617 & 493.0  &  0.000004  &  59076.470079 & 512.5  &  0.000006 \\
58342.241178 & 11.0   &  0.000004  &  59049.383671 & 494.0  &  0.000004  &  59077.201896 & 513.0  &  0.000004 \\
58342.973078 & 11.5   &  0.000004  &  59050.116215 & 494.5  &  0.000006  &  59077.934281 & 513.5  &  0.000006 \\
58343.705279 & 12.0   &  0.000004  &  59050.847785 & 495.0  &  0.000004  &  59078.666081 & 514.0  &  0.000005 \\
58344.437009 & 12.5   &  0.000006  &  59051.580292 & 495.5  &  0.000006  &  59079.398485 & 514.5  &  0.000007 \\
58345.169391 & 13.0   &  0.000005  &  59052.311934 & 496.0  &  0.000005  &  59080.130262 & 515.0  &  0.000005 \\
58345.901131 & 13.5   &  0.000006  &  59053.044348 & 496.5  &  0.000007  &  59080.862593 & 515.5  &  0.000006 \\
58346.633499 & 14.0   &  0.000005  &  59053.776090 & 497.0  &  0.000006  &  59081.594352 & 516.0  &  0.000004 \\
58347.365392 & 14.5   &  0.000007  &  59054.508387 & 497.5  &  0.000006  &  59082.326656 & 516.5  &  0.000005 \\
58348.097205 & 15.0   &  0.000074  &  59055.240325 & 498.0  &  0.000005  &  59083.058456 & 517.0  &  0.000004 \\
58348.829518 & 15.5   &  0.000074  &  59055.972499 & 498.5  &  0.000004  &  59083.790713 & 517.5  &  0.000004 \\
\hline
\end{tabular}

\textit{Notes.} Integer and half-integer cycle numbers, as above, refer to primary and secondary eclipses, respectively.   
\end{table*}

\begin{table*}
\caption{Eclipse times of BG Ind binary B determined from \textit{TESS} observations.}
 \label{tab:BGIndB_etv_TESS}
\begin{tabular}{@{}lrllrllrl}
\hline
BJD & Cycle  & std. dev. & BJD & Cycle  & std. dev. & BJD & Cycle  & std. dev. \\ 
$-2\,400\,000$ & no. &   \multicolumn{1}{c}{$(d)$} & $-2\,400\,000$ & no. &   \multicolumn{1}{c}{$(d)$} & $-2\,400\,000$ & no. &   \multicolumn{1}{c}{$(d)$} \\ 
\hline
58325.543752 & -0.5 &  0.000062 & 58337.430905 & 22.0 &  0.000054 & 58350.902971 & 47.5 &  0.000085 \\
58325.807997 & 0.0 &   0.000046 & 58337.693064 & 22.5 &  0.000104 & 58351.167488 & 48.0 &  0.000079 \\
58326.072387 & 0.5 &   0.000072 & 58337.959793 & 23.0 &  0.000049 & 58351.431071 & 48.5 &  0.000088 \\
58326.336575 & 1.0 &   0.000064 & 58338.222995 & 23.5 &  0.000055 & 58351.696086 & 49.0 &  0.000054 \\
58326.600567 & 1.5 &   0.000072 & 58338.487016 & 24.0 &  0.000049 & 58351.959199 & 49.5 &  0.000097 \\
58326.864689 & 2.0 &   0.000053 & 58339.808135 & 26.5 &  0.000070 & 58352.224055 & 50.0 &  0.000051 \\
58327.128861 & 2.5 &   0.000066 & 58340.072263 & 27.0 &  0.000055 & 58352.488050 & 50.5 &  0.000058 \\
58327.392666 & 3.0 &   0.000052 & 58340.336199 & 27.5 &  0.000079 & 58352.752014 & 51.0 &  0.000051 \\
58327.657142 & 3.5 &   0.000077 & 58340.601135 & 28.0 &  0.000059 & 58353.016148 & 51.5 &  0.000071 \\
58327.921372 & 4.0 &   0.000046 & 58340.864891 & 28.5 &  0.000102 & 59036.438310 & 1345.0 &  0.000073 \\
58328.185258 & 4.5 &   0.000060 & 58341.129590 & 29.0 &  0.000056 & 59036.702049 & 1345.5 &  0.000075 \\
58328.449821 & 5.0 &   0.000058 & 58341.392795 & 29.5 &  0.000076 & 59036.966367 & 1346.0 &  0.000071 \\
58328.713709 & 5.5 &   0.000060 & 58341.657591 & 30.0 &  0.000068 & 59037.230717 & 1346.5 &  0.000068 \\
58328.977792 & 6.0 &   0.000048 & 58341.921373 & 30.5 &  0.000092 & 59037.494934 & 1347.0 &  0.000061 \\
58329.241446 & 6.5 &   0.000062 & 58342.185766 & 31.0 &  0.000078 & 59037.758430 & 1347.5 &  0.000073 \\
58329.506583 & 7.0 &   0.000058 & 58342.450293 & 31.5 &  0.000057 & 59038.023264 & 1348.0 &  0.000068 \\
58329.770573 & 7.5 &   0.000072 & 58342.714094 & 32.0 &  0.000053 & 59038.286595 & 1348.5 &  0.000054 \\
58330.034386 & 8.0 &   0.000069 & 58342.978300 & 32.5 &  0.000077 & 59038.551986 & 1349.0 &  0.000089 \\
58330.298768 & 8.5 &   0.000102 & 58343.242670 & 33.0 &  0.000051 & 59038.815125 & 1349.5 &  0.000088 \\
58330.562823 & 9.0 &   0.000064 & 58343.506139 & 33.5 &  0.000082 & 59039.080123 & 1350.0 &  0.000068 \\
58330.826710 & 9.5 &   0.000060 & 58343.770936 & 34.0 &  0.000070 & 59039.342745 & 1350.5 &  0.000095 \\
58331.091287 & 10.0 &  0.000051 & 58344.034952 & 34.5 &  0.000081 & 59039.608385 & 1351.0 &  0.000062 \\
58331.355655 & 10.5 &  0.000094 & 58344.299072 & 35.0 &  0.000080 & 59039.871119 & 1351.5 &  0.000082 \\
58331.619925 & 11.0 &  0.000054 & 58344.562888 & 35.5 &  0.000099 & 59040.137118 & 1352.0 &  0.000074 \\
58331.883907 & 11.5 &  0.000072 & 58344.827200 & 36.0 &  0.000046 & 59040.400668 & 1352.5 &  0.000074 \\
58332.147667 & 12.0 &  0.000050 & 58345.091215 & 36.5 &  0.000073 & 59040.665367 & 1353.0 &  0.000086 \\
58332.411725 & 12.5 &  0.000079 & 58345.355528 & 37.0 &  0.000049 & 59040.928474 & 1353.5 &  0.000086 \\
58332.676452 & 13.0 &  0.000059 & 58345.619873 & 37.5 &  0.000074 & 59041.193535 & 1354.0 &  0.000075 \\
58332.940235 & 13.5 &  0.000058 & 58345.884337 & 38.0 &  0.000053 & 59041.456972 & 1354.5 &  0.000074 \\
58333.204648 & 14.0 &  0.000046 & 58346.148236 & 38.5 &  0.000069 & 59041.721550 & 1355.0 &  0.000070 \\
58333.468688 & 14.5 &  0.000074 & 58346.412325 & 39.0 &  0.000057 & 59041.985219 & 1355.5 &  0.000102 \\
58333.732522 & 15.0 &  0.000044 & 58346.676294 & 39.5 &  0.000100 & 59042.249607 & 1356.0 &  0.000060 \\
58333.997359 & 15.5 &  0.000065 & 58346.940771 & 40.0 &  0.000071 & 59042.513262 & 1356.5 &  0.000073 \\
58334.261381 & 16.0 &  0.000069 & 58347.204409 & 40.5 &  0.000097 & 59042.777673 & 1357.0 &  0.000064 \\
58334.525599 & 16.5 &  0.000072 & 58347.468581 & 41.0 &  0.000263 & 59043.041442 & 1357.5 &  0.000112 \\
58334.789683 & 17.0 &  0.000058 & 58347.736341 & 41.5 &  0.000426 & 59043.306328 & 1358.0 &  0.000066 \\
58335.053444 & 17.5 &  0.000070 & 58347.997914 & 42.0 &  0.000197 & 59043.570070 & 1358.5 &  0.000101 \\
58335.317858 & 18.0 &  0.000069 & 58348.526424 & 43.0 &  0.000392 & 59043.834600 & 1359.0 &  0.000077 \\
58335.581721 & 18.5 &  0.000080 & 58349.053670 & 44.0 &  0.000225 & 59044.098701 & 1359.5 &  0.000065 \\
58335.846166 & 19.0 &  0.000048 & 58349.316776 & 44.5 &  0.000139 & 59044.362603 & 1360.0 &  0.000078 \\
58336.110542 & 19.5 &  0.000061 & 58349.582154 & 45.0 &  0.000055 & 59044.626253 & 1360.5 &  0.000079 \\
58336.374765 & 20.0 &  0.000048 & 58349.845864 & 45.5 &  0.000060 & 59044.891275 & 1361.0 &  0.000081 \\
58336.638703 & 20.5 &  0.000079 & 58350.110733 & 46.0 &  0.000050 & 59045.155782 & 1361.5 &  0.000080 \\
58336.902547 & 21.0 &  0.000061 & 58350.374797 & 46.5 &  0.000088 & 59045.419161 & 1362.0 &  0.000068 \\
58337.166804 & 21.5 &  0.000084 & 58350.639299 & 47.0 &  0.000067 & 59045.683175 & 1362.5 &  0.000077 \\
\hline
\end{tabular}

\textit{Notes.} Integer and half-integer cycle numbers, as above, refer to primary and secondary eclipses, respectively.   
\end{table*}

\addtocounter{table}{-1}
\begin{table*}
\caption{continued ({\em TESS} binary B eclipse times)}
\begin{tabular}{@{}lrllrllrl}
\hline
BJD & Cycle  & std. dev. & BJD & Cycle  & std. dev. & BJD & Cycle  & std. dev. \\ 
$-2\,400\,000$ & no. &   \multicolumn{1}{c}{$(d)$} & $-2\,400\,000$ & no. &   \multicolumn{1}{c}{$(d)$} & $-2\,400\,000$ & no. &   \multicolumn{1}{c}{$(d)$} \\ 
\hline
59045.946568 & 1363.0 &  0.000077  & 59058.099423 & 1386.0 &  0.000073  & 59070.513321 & 1409.5 &  0.000065 \\
59046.210665 & 1363.5 &  0.000093  & 59058.362578 & 1386.5 &  0.000088  & 59070.778828 & 1410.0 &  0.000054 \\
59046.476190 & 1364.0 &  0.000080  & 59058.627701 & 1387.0 &  0.000082  & 59071.042587 & 1410.5 &  0.000075 \\
59046.739114 & 1364.5 &  0.000086  & 59058.891008 & 1387.5 &  0.000076  & 59071.306873 & 1411.0 &  0.000055 \\
59047.003837 & 1365.0 &  0.000072  & 59059.156332 & 1388.0 &  0.000087  & 59075.269512 & 1418.5 &  0.000073 \\
59047.267302 & 1365.5 &  0.000078  & 59059.419938 & 1388.5 &  0.000098  & 59075.534009 & 1419.0 &  0.000081 \\
59047.532523 & 1366.0 &  0.000084  & 59059.684201 & 1389.0 &  0.000085  & 59075.797714 & 1419.5 &  0.000062 \\
59047.795808 & 1366.5 &  0.000088  & 59059.947683 & 1389.5 &  0.000082  & 59076.061853 & 1420.0 &  0.000062 \\
59048.060856 & 1367.0 &  0.000103  & 59060.212760 & 1390.0 &  0.000092  & 59076.326308 & 1420.5 &  0.000064 \\
59049.380015 & 1369.5 &  0.000075  & 59060.475958 & 1390.5 &  0.000095  & 59076.590032 & 1421.0 &  0.000062 \\
59049.645507 & 1370.0 &  0.000053  & 59062.059702 & 1393.5 &  0.000070  & 59076.853883 & 1421.5 &  0.000055 \\
59049.908823 & 1370.5 &  0.000083  & 59062.324947 & 1394.0 &  0.000050  & 59077.118281 & 1422.0 &  0.000062 \\
59050.174129 & 1371.0 &  0.000064  & 59062.588783 & 1394.5 &  0.000071  & 59077.382065 & 1422.5 &  0.000067 \\
59050.436919 & 1371.5 &  0.000082  & 59062.853569 & 1395.0 &  0.000059  & 59077.646390 & 1423.0 &  0.000059 \\
59050.701683 & 1372.0 &  0.000067  & 59063.117122 & 1395.5 &  0.000067  & 59077.910784 & 1423.5 &  0.000066 \\
59050.964997 & 1372.5 &  0.000087  & 59063.381628 & 1396.0 &  0.000061  & 59078.174122 & 1424.0 &  0.000060 \\
59051.230902 & 1373.0 &  0.000069  & 59063.645362 & 1396.5 &  0.000089  & 59078.438383 & 1424.5 &  0.000070 \\
59051.494492 & 1373.5 &  0.000072  & 59063.910690 & 1397.0 &  0.000048  & 59078.702977 & 1425.0 &  0.000068 \\
59051.759397 & 1374.0 &  0.000084  & 59064.174613 & 1397.5 &  0.000075  & 59078.966824 & 1425.5 &  0.000072 \\
59052.023123 & 1374.5 &  0.000066  & 59064.438258 & 1398.0 &  0.000061  & 59079.231510 & 1426.0 &  0.000067 \\
59052.286863 & 1375.0 &  0.000092  & 59064.701572 & 1398.5 &  0.000073  & 59079.494570 & 1426.5 &  0.000069 \\
59052.550981 & 1375.5 &  0.000093  & 59064.966169 & 1399.0 &  0.000058  & 59079.759833 & 1427.0 &  0.000052 \\
59052.815849 & 1376.0 &  0.000072  & 59065.229515 & 1399.5 &  0.000056  & 59080.023589 & 1427.5 &  0.000084 \\
59053.079435 & 1376.5 &  0.000067  & 59065.494779 & 1400.0 &  0.000064  & 59080.288840 & 1428.0 &  0.000071 \\
59053.343861 & 1377.0 &  0.000073  & 59065.758511 & 1400.5 &  0.000068  & 59080.550777 & 1428.5 &  0.000060 \\
59053.607858 & 1377.5 &  0.000073  & 59066.022631 & 1401.0 &  0.000049  & 59080.816015 & 1429.0 &  0.000081 \\
59053.871578 & 1378.0 &  0.000075  & 59066.286854 & 1401.5 &  0.000080  & 59081.078603 & 1429.5 &  0.000104 \\
59054.135531 & 1378.5 &  0.000086  & 59066.550834 & 1402.0 &  0.000066  & 59081.343803 & 1430.0 &  0.000056 \\
59054.400856 & 1379.0 &  0.000070  & 59066.814401 & 1402.5 &  0.000071  & 59081.607927 & 1430.5 &  0.000072 \\
59054.664783 & 1379.5 &  0.000076  & 59067.079190 & 1403.0 &  0.000075  & 59081.872450 & 1431.0 &  0.000053 \\
59054.929447 & 1380.0 &  0.000116  & 59067.343567 & 1403.5 &  0.000082  & 59082.135381 & 1431.5 &  0.000083 \\
59055.192947 & 1380.5 &  0.000070  & 59067.608362 & 1404.0 &  0.000055  & 59082.401060 & 1432.0 &  0.000060 \\
59055.457163 & 1381.0 &  0.000070  & 59067.872493 & 1404.5 &  0.000062  & 59082.663637 & 1432.5 &  0.000064 \\
59055.721318 & 1381.5 &  0.000065  & 59068.136789 & 1405.0 &  0.000065  & 59082.928962 & 1433.0 &  0.000067 \\
59055.985884 & 1382.0 &  0.000059  & 59068.399954 & 1405.5 &  0.000074  & 59083.192191 & 1433.5 &  0.000088 \\
59056.250243 & 1382.5 &  0.000079  & 59068.665304 & 1406.0 &  0.000058  & 59083.457546 & 1434.0 &  0.000064 \\
59056.514623 & 1383.0 &  0.000070  & 59068.928573 & 1406.5 &  0.000051  & 59083.720955 & 1434.5 &  0.000100 \\
59056.778127 & 1383.5 &  0.000087  & 59069.193428 & 1407.0 &  0.000064  & 59083.985845 & 1435.0 &  0.000051 \\
59057.043001 & 1384.0 &  0.000055  & 59069.456733 & 1407.5 &  0.000063  & 59084.249018 & 1435.5 &  0.000080 \\
59057.306395 & 1384.5 &  0.000079  & 59069.721349 & 1408.0 &  0.000069  & 59084.513792 & 1436.0 &  0.000065 \\
59057.570656 & 1385.0 &  0.000075  & 59069.984985 & 1408.5 &  0.000060  &&& \\
59057.834665 & 1385.5 &  0.000084  & 59070.250315 & 1409.0 &  0.000075  &&&  \\
\hline
\end{tabular}
\end{table*}



\begin{table*}
\caption{Eclipse times of BG Ind binary A determined from archival, ground-based photometric measurements.}
 \label{tab:BGIndA_etv_ground}
\begin{tabular}{@{}lrllrllrl}
\hline
BJD & Cycle  & std. dev. & BJD & Cycle  & std. dev. & BJD & Cycle  & std. dev. \\ 
$-2\,400\,000$ & no. &   \multicolumn{1}{c}{$(d)$} & $-2\,400\,000$ & no. &   \multicolumn{1}{c}{$(d)$} & $-2\,400\,000$ & no. &   \multicolumn{1}{c}{$(d)$} \\ 
\hline
45905.747600$^a$ & -8483.5  &  0.001000  &  56203.239825 & -1450.0  &  0.000915  &  56574.380595 & -1196.5  &  0.000149 \\
46670.720391$^b$ & -7961.0  &  0.001007  &  56205.439280 & -1448.5  &  0.000161  &  56577.309551 & -1194.5  &  0.000229 \\
46695.610663$^b$ & -7944.0  &  0.001082  &  56211.296461 & -1444.5  &  0.000117  &  56585.361049 & -1189.0  &  0.000328 \\
50749.603297$^c$ & -5175.0  &  0.000200  &  56213.491330 & -1443.0  &  0.000201  &  56599.268820 & -1179.5  &  0.000453 \\
50752.535697$^c$ & -5173.0  &  0.001000  &  56219.346595 & -1439.0  &  0.000305  &  56601.468138 & -1178.0  &  0.000489 \\
50760.584197$^c$ & -5167.5  &  0.000200  &  56241.309142 & -1424.0  &  0.000095  &  56604.395227 & -1176.0  &  0.000115 \\
51096.587705$^c$ & -4938.0  &  0.000200  &  56246.434520 & -1420.5  &  0.000777  &  56607.321861 & -1174.0  &  0.000190 \\
51137.582236$^c$ & -4910.0  &  0.000200  &  56249.362519 & -1418.5  &  0.000130  &  56615.374282 & -1168.5  &  0.000208 \\
55783.058458$^d$ & -1737.0  &  0.000800  &  56252.292900 & -1416.5  &  0.000312  &  56819.609731 & -1029.0  &  0.000231 \\
55837.954960$^d$ & -1699.5  &  0.000200  &  56257.416438 & -1413.0  &  0.000460  &  56822.538145 & -1027.0  &  0.000935 \\
56114.664682 & -1510.5  &  0.000263  &  56260.343245 & -1411.0  &  0.000176  &  56868.660034 & -995.5   &  0.000387 \\
56120.518954 & -1506.5  &  0.000139  &  56450.673609 & -1281.0  &  0.000270  &  56871.584259 & -993.5   &  0.000166 \\
56125.644943 & -1503.0  &  0.000088  &  56453.601199 & -1279.0  &  0.000060  &  56888.423844 & -982.0   &  0.000073 \\
56128.573285 & -1501.0  &  0.000546  &  56464.581939 & -1271.5  &  0.000063  &  56893.547481 & -978.5   &  0.000130 \\
56134.428810 & -1497.0  &  0.000679  &  56480.684869 & -1260.5  &  0.000345  &  56902.333131 & -972.5   &  0.000358 \\
56139.552054 & -1493.5  &  0.000130  &  56483.612309 & -1258.5  &  0.000133  &  56904.529609 & -971.0   &  0.000119 \\
56150.534786 & -1486.0  &  0.000459  &  56491.664083 & -1253.0  &  0.000358  &  56907.456133 & -969.0   &  0.000216 \\
56153.464094 & -1484.0  &  0.000251  &  56500.449720 & -1247.0  &  0.000224  &  56913.311888 & -965.0   &  0.000214 \\
56158.587494 & -1480.5  &  0.000071  &  56505.572608 & -1243.5  &  0.000104  &  56923.563562 & -958.0   &  0.000627 \\
56161.515700 & -1478.5  &  0.000094  &  56508.501528 & -1241.5  &  0.000094  &  56926.489918 & -956.0   &  0.000094 \\
56166.641916 & -1475.0  &  0.000295  &  56524.605414 & -1230.5  &  0.000128  &  56934.541525 & -950.5   &  0.000231 \\
56167.373624 & -1474.5  &  0.000103  &  56538.511594 & -1221.0  &  0.000128  &  56935.274575 & -950.0   &  0.000141 \\
56175.425003 & -1469.0  &  0.000060  &  56541.440387 & -1219.0  &  0.000123  &  56945.521544 & -943.0   &  0.000392 \\
56177.620729 & -1467.5  &  0.025388  &  56547.296407 & -1215.0  &  0.000896  &  56948.450675 & -941.0   &  0.000107 \\
56178.352518 & -1467.0  &  0.000088  &  56557.543911 & -1208.0  &  0.000203  &  56951.379494 & -939.0   &  0.000137 \\
56186.407006 & -1461.5  &  0.000160  &  56558.279472 & -1207.5  &  0.000209  &  56954.308374 & -937.0   &  0.000156 \\
56188.603921 & -1460.0  &  0.000307  &  56560.472694 & -1206.0  &  0.000120  &  56967.484429 & -928.0   &  0.000531 \\
56191.527603 & -1458.0  &  0.000357  &  56565.594480 & -1202.5  &  0.001205  &  56970.412294 & -926.0   &  0.000236 \\
56197.385670 & -1454.0  &  0.000289  &  56566.330280 & -1202.0  &  0.000089  &               &          &           \\
56202.511063 & -1450.5  &  0.000278  &  56569.257382 & -1200.0  &  0.000281  &               &          &           \\   
\hline
\end{tabular}

\textit{Notes.} Integer and half-integer cycle numbers refer to primary and secondary eclipses, respectively. Times of minima between cycle nos. $-1737.0$ and $-926.0$ were determined from WASP measurements. The sources of the few other, older eclipse times are follows: (a) \citet{vanhammemanfroid88}; (b) present paper, determined from the ESO archival timeseries, see Sect.~\ref{subsect:lc_ESO_archive}; (c) present paper, determined from unpublished observations of Jens Viggo Clausen; (d) present paper, observations of co-author MB.

\end{table*}

\begin{figure*}
\begin{center}
\includegraphics[width=0.49 \textwidth]{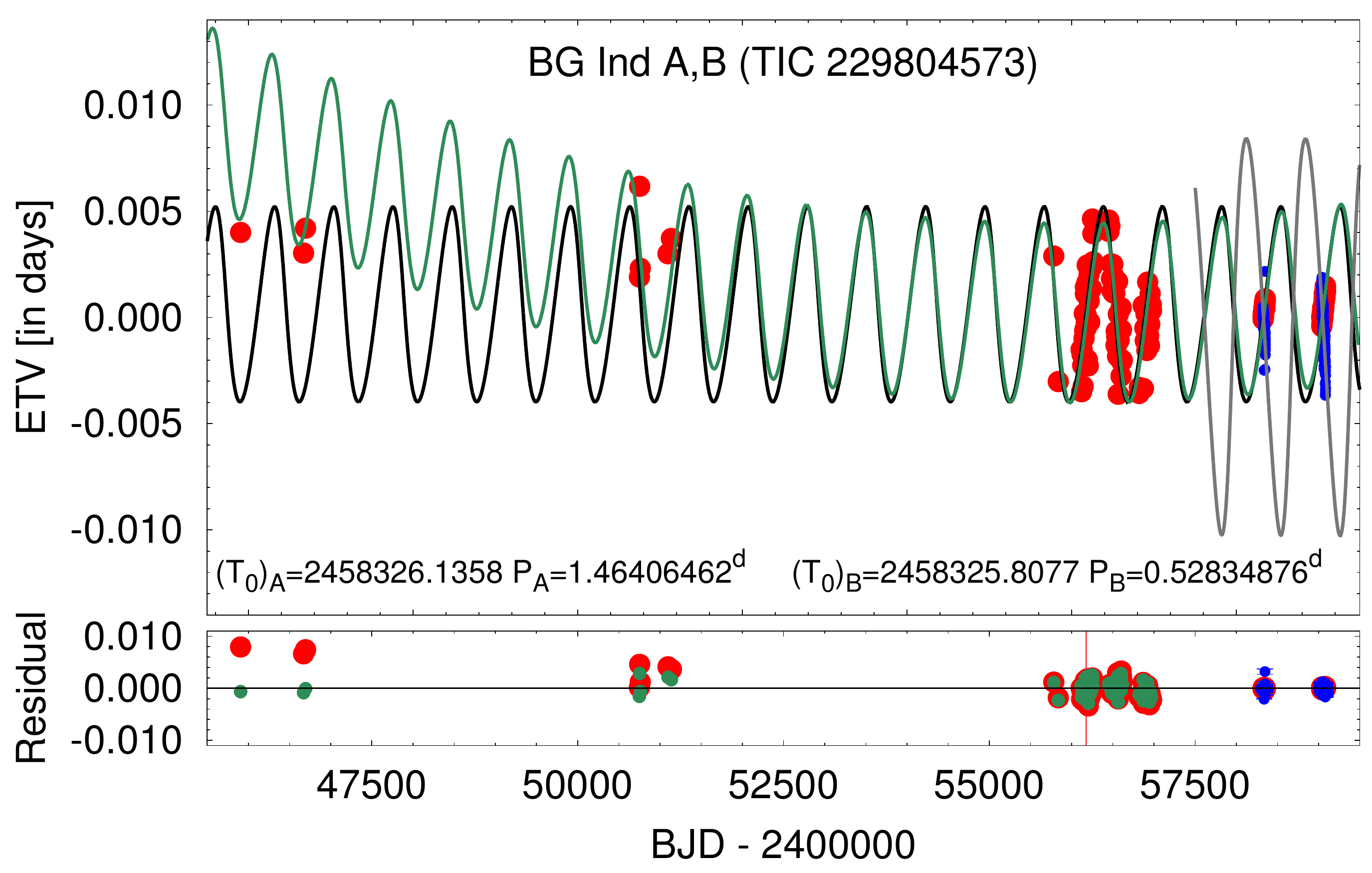}\includegraphics[width=0.49 \textwidth]{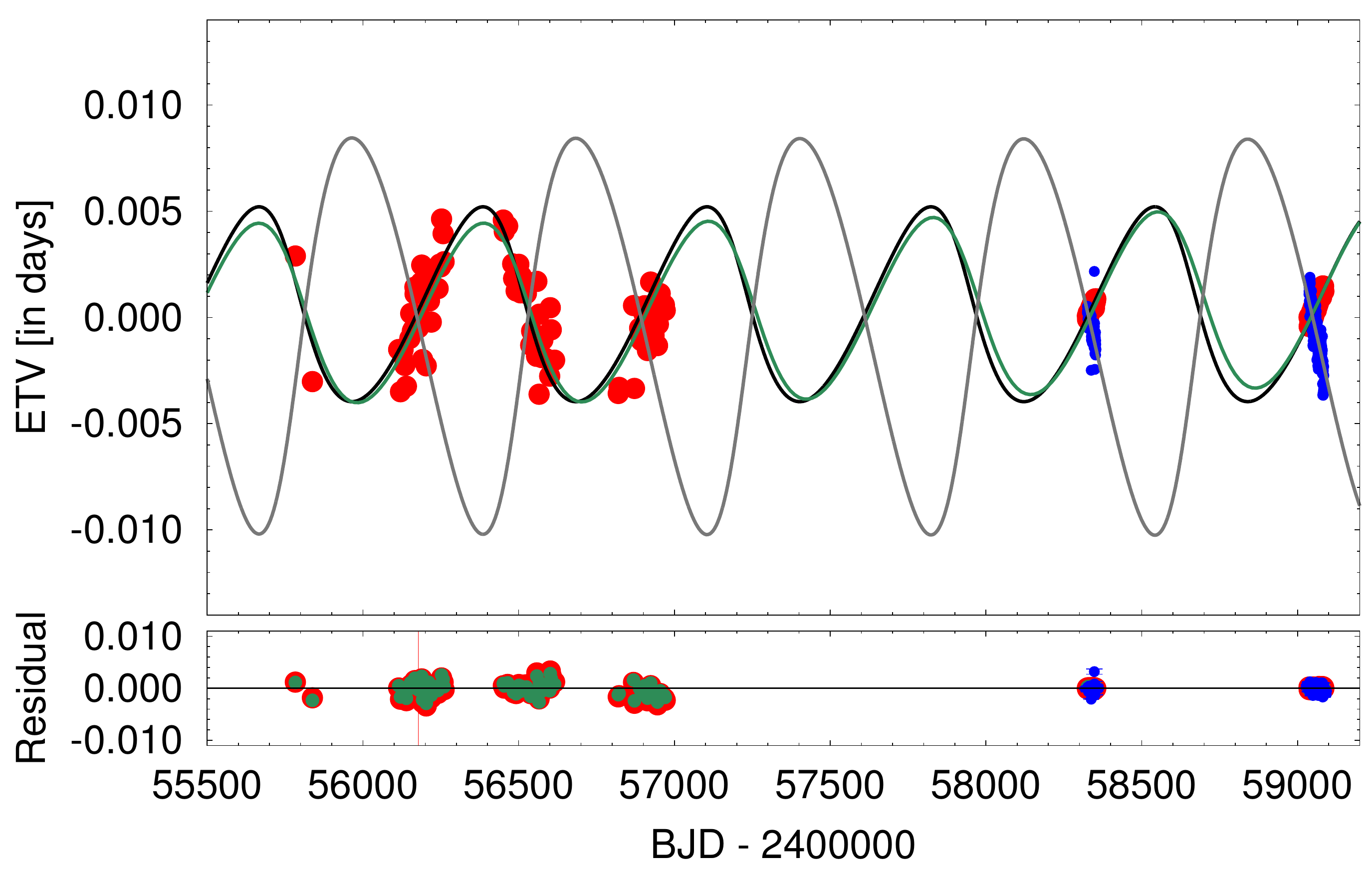}
\caption{Eclipse timing variations of BG\,Ind. \textit{Left panel} shows all the available observations, while in the \textit{right panel} we zoom in on the regions of the better-covered WASP and \textit{TESS} data. Larger red circles represent ETV points calculated from the observed eclipse events of binary A, while the smaller blue circles stand for the ETVs of binary B. Note, for simplicity, we do not separate primary and secondary eclipses. (The validity of this can easily be verified since both binaries have circular orbits and, furthermore, the primary and secondary eclipses within each binary can be calculated with the same accuracy due to their similar depths.) Black and grey lines stand for the combined spectro-photodynamical model ETV solution (Sect.~\ref{Sect:photodyn}) for binary A and B, respectively, while the green line denotes the preliminary, `classic', analytic LTTE+quadratic ETV solution discussed in Sect.~\ref{sect:classicETV}. (Note, for clarity, in the left panel, the ETV solution of binary B, i.e., the grey curve, is plotted only for the narrow interval around the \textit{TESS} observations.) The residuals of the observed vs. modelled ETVs are plotted in the bottom panel. Here, as above, red and blue dots represent the residuals of binary A and B ETV points against the spectro-photodynamical model, while green dots stand for the residuals of binary A data against the analytic LTTE+quadratic ETV model.} 
\label{fig:etvs} 
\end{center}
\end{figure*}  

\subsubsection{BG Ind ETV Results}
\label{sect:classicETV}

The overall ETV curves for BG\,Ind A and B are plotted in Fig.~\ref{fig:etvs} along with the best-fit spectro-photodynamical model that is described in the next section. The ETV curve of BG Ind A exhibits a clearly cyclic pattern with a period of $\sim2$~years. Even in the absence of any other indications of additional stars in the system, the most plausible explanation of this ETV behavior would be the light-travel time effect (LTTE) caused by a gravitationally bound, distant, third component. Therefore, we carried out, a preliminary, `traditional' analysis of the ETV curves of binary A by fitting the LTTE-term with our analytic ETV-solver \citep{borkoetal15}. We found that the very first eight ETV points deviate systematically from the LTTE solution. Therefore, we added a quadratic term to the analysis and obtained the following quadratic ephemeris:
\begin{eqnarray}
T_\mathrm{pri}&=&2\,458\,326.13561(7)+1.46406518(8)\times E \nonumber \\
&&+1.8(3)\times10^{-10}\times E^2.
\label{eq:ETV_quad}
\end{eqnarray} 
We also tabulate the parameters of this preliminary LTTE solution in Table~\ref{tab:ETV_lite}, and plot this simple model together with the spectro-photodynamical model, in Fig.~\ref{fig:etvs}.

\begin{table}
\centering
\caption{Light-travel-time orbital solution for BG~Ind~A from a classical ETV analysis of its outer orbit}
 \label{tab:ETV_lite}
\begin{tabular}{@{}ll}
\hline
Parameter & value \\
\hline
$P_\mathrm{out}$ [d] & $721.2\pm0.9$ \\ 
$a_\mathrm{AB}\sin i_\mathrm{out}$ [R$_\odot$] & $161\pm9$ \\ 
$e_\mathrm{out}$ & $0.21\pm0.06$ \\
$\omega_\mathrm{out}$ [$\degr$] & $345\pm8$ \\
$\tau_\mathrm{out}$ [BJD] & $2\,458\,678\pm16$ \\
$f(m_\mathrm{B})$ [M$_\odot$] & $0.11\pm0.02$ \\
$K_\mathrm{A}$ [km\,s$^{-1}$] & $11.5\pm0.6$ \\
$\dot P_\mathrm{A}/P_\mathrm{A}$ [$10^{-8}$\,yr$^{-1}$] & $6.1\pm 0.4$\\
\hline
\end{tabular}

\textit{Notes.} $a_\mathrm{AB}\sin i_\mathrm{out}$ denotes the line-of-sight projected semi-major axis of the outer orbit of binary A around the center of mass of the quadruple system, while the other orbital elements and associated parameters are noted in their usual manner. Moreover, we tabulate two derived parameters: $f(m_\mathrm{B})$, and $K_\mathrm{A}$ which are the mass function and the amplitude of the radial velocity curve of the center of mass of binary A on its outer orbit. Finally, in the last row, we give the rate of the continuous period variation of binary A, which is derived from eqn.~(\ref{eq:ETV_quad}). 
\end{table}

Turning to the ETV points of binary B, they appear to be moving with the opposite phase to that of the bright binary A which makes it very likely that the two binaries form a bound, quite tight quadruple system. As we will discuss below in Sects.~\ref{Sect:photodyn} and \ref{sec:discussion} our detailed analysis robustly confirms this hypothesis.


\subsection{Radial Velocity data}
\label{subsect_RVdata}

\begin{figure}
\begin{center}
\includegraphics[width=0.47 \textwidth]{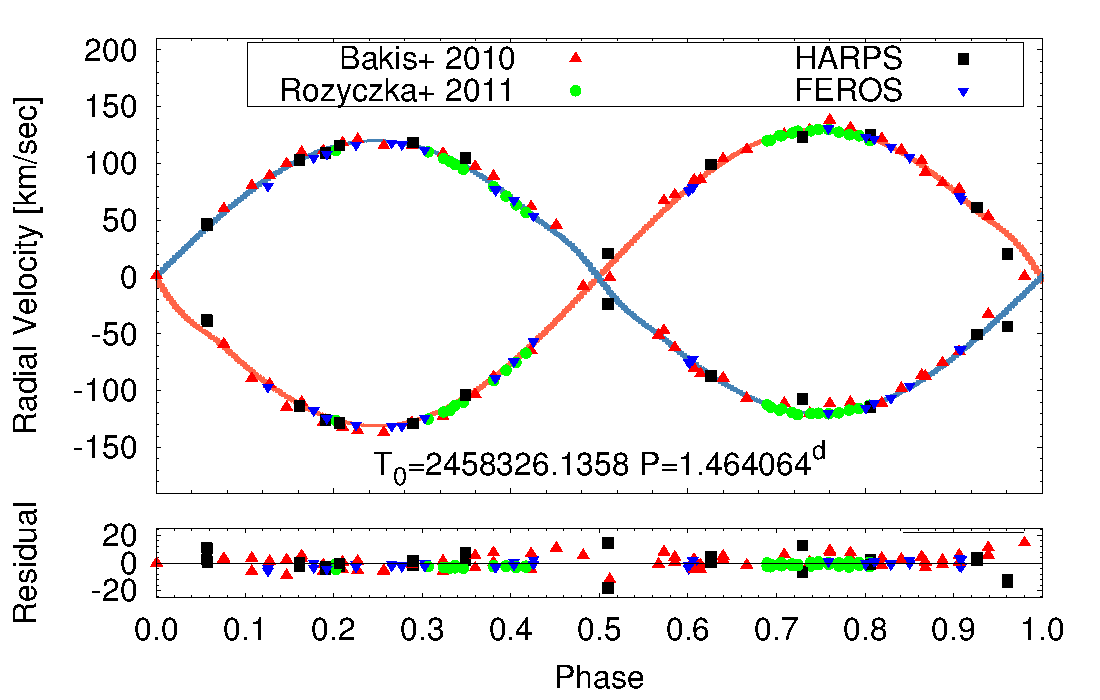}
\caption{Phase-folded RV curve of the brighter binary, A, in BG Ind after the removal of the contribution of the orbital motion around the center of mass of the quadruple system, and the systemic RV ($\gamma$) of the quadruple, as well. (The values to be removed were calculated from the best-fit joint spectro-photodynamical model, see below in Sect.~\ref{Sect:photodyn}). The origin of each set of data points are noted in the key box. Red and blue lines stand for the model solutions for the photometric primary and secondary (spectroscopic secondary and primary) components, respectively.} 
\label{fig:RVAab_fold} 
\end{center}
\end{figure}  

We used three sets of radial velocity (RV) data for our analysis. These are as follows: $(i)$ \citet{bakisetal10} have obtained 41 RVs between JDs 2\,453\,968 and 2\,453\,996 (i.e. in August 2006) with the High Efficiency and Resolution Canterbury University Large Echelle Spectrograph (HERCULES) of the Department of Physics and Astronomy, New Zealand; $(ii)$ an additional 23 RVs for both components of binary A between JDs 2\,454\,363 and 2\,454\,376 (September/October 2007) were collected by \citet{rozyczkaetal11} with the fibre-fed Giraffe spectrograph on the 1.9-m Radcliffe telescope at the South African Astronomical Observatory; and $(iii)$ finally, we found in the ESO publicly accessible archive\footnote{\url{http://archive.eso.org/cms.html}} a number of spectra taken with the HARPS and FEROS spectrographs between JDs 2\,453\,191 and 2\,456\,910. From these data we determined an additional 54 RV points of both components of binary A. 

To measure the RVs from these latter spectra we used the broadening function method \citep{rucinski92} as implemented in the software package {\sc RaveSpan} \citep{pileckietal17}. The template used for the analysis was a synthetic spectrum for a star with $T_\mathrm{eff}=7000$\,K, $\log g=3.5$. We used a simultaneous least-squares fit of two rotationally-broadened profiles to measure the radial velocities of the two stars from the broadening profile. The RV points obtained in this way are tabulated in Table~\ref{tab:HARPSFEROS}.

The phase-folded RV points (after the correcting for the orbital motion around the center of mass of the whole quadruple system) together with the best-fit photodynamical solution (see below, in Sect.~\ref{Sect:photodyn}) are plotted in Fig.~\ref{fig:RVAab_fold}.

\begin{table}
\caption{Unpublished ESO archive RV data for BG Ind.}
 \label{tab:HARPSFEROS}
\begin{tabular}{@{}lrrrrr}
\hline
BJD & $V_\mathrm{Aa}$  & $\sigma_\mathrm{Aa}$ & $V_\mathrm{Ab}$  & $\sigma_\mathrm{Ab}$ & Instr.\\ 
$-2\,400\,000$ & \multicolumn{4}{c}{~~~~(km s$^{-1})$} \\ 
\hline
53191.746239 &  104.898 &  2.107 &   17.981 &  2.731 &  HARPS \\
53191.747289 &  103.736 &  2.558 &   16.724 &  2.676 &  HARPS \\
53196.615059 &  134.576 &  2.113 &  -30.806 &  1.888 &  FEROS \\
53196.616399 &  133.978 &  2.006 &  -30.566 &  1.895 &  FEROS \\
53205.719319 &  -16.673 &  2.060 &  133.718 &  1.792 &  FEROS \\
53205.720649 &  -17.291 &  2.040 &  134.728 &  1.842 &  FEROS \\
55468.519001 &  161.283 &  2.291 &  -54.899 &  2.114 &  HARPS \\
55471.513061 &  173.835 &  2.285 &  -70.150 &  1.922 &  HARPS \\
55477.491091 &  175.936 &  2.223 &  -70.213 &  1.905 &  HARPS \\
55478.473551 &   14.546 &  2.592 &   85.063 &  1.895 &  HARPS \\
55479.598661 &  -48.246 &  1.202 &  181.327 &  2.374 &  HARPS \\
55479.711481 &  -55.896 &  2.364 &  183.279 &  1.943 &  HARPS \\
55535.522132 &    5.406 &  2.841 &  117.879 &  2.621 &  HARPS \\
56449.951656 &   14.266 &  1.994 &   60.955 &  1.400 &  HARPS \\
56450.949046 &  148.986 &  2.185 &  -85.547 &  1.943 &  HARPS \\
56473.803176 &  -77.384 &  2.044 &  160.731 &  1.827 &  FEROS \\
56473.844686 &  -69.309 &  2.033 &  151.984 &  1.870 &  FEROS \\
56475.840366 &  145.259 &  2.001 &  -87.001 &  1.924 &  FEROS \\
56475.964466 &  153.088 &  2.046 &  -93.602 &  1.869 &  FEROS \\
56553.665337 &  138.067 &  2.543 &  -70.208 &  2.350 &  HARPS \\
56577.496218 &  -51.494 &  2.467 &  133.971 &  1.974 &  HARPS \\
56906.616809 &  112.284 &  2.517 &    2.106 &  2.109 &  FEROS \\
56908.568369 &  -60.329 &  1.962 &  189.225 &  1.840 &  FEROS \\
56908.644099 &  -52.057 &  2.098 &  180.060 &  1.812 &  FEROS \\
56908.702749 &  -37.210 &  1.950 &  164.030 &  1.893 &  FEROS \\
56908.784749 &   -5.251 &  2.000 &  129.609 &  1.942 &  FEROS \\
56908.787889 &   -5.126 &  1.842 &  127.316 &  1.864 &  FEROS \\
56909.512639 &  125.816 &  2.033 &  -15.650 &  1.842 &  FEROS \\
56909.808189 &  -13.712 &  1.987 &  137.056 &  1.815 &  FEROS \\
56910.570019 &  138.822 &  1.930 &  -38.326 &  1.623 &  FEROS \\
56910.645749 &  163.451 &  2.081 &  -58.828 &  1.893 &  FEROS \\
56910.715679 &  174.478 &  2.007 &  -71.379 &  1.874 &  FEROS \\
56910.774629 &  175.871 &  2.132 &  -72.403 &  1.874 &  FEROS \\
56910.829279 &  169.785 &  2.038 &  -65.114 &  1.846 &  FEROS \\
\hline
\end{tabular}
\end{table}

\section{Joint analysis of the available data}
\label{Sect:photodyn}

We used the software package {\sc Lightcurvefactory} \citep[see][and further references therein]{borkoetal19,borkoetal20} to carry out a complex spectro-photodynamical modeling of the system based on the data collected in Sect.~\ref{sec:obsdata}. {\sc Lightcurvefactory} calculates stellar positions and velocities for each object and emulates the light-, RV and ETV curves of any arbitrary quadruple system (having either 2+2 or 2+1+1 hierarchies), including mutual eclipses amongst any two (or more) components. Moreover, the software may (optionally) use built-in, pre-calculated \texttt{PARSEC} isochrone tables\footnote{These tables generated via the web based tool CMD 3.3,\\  {\url{http://stev.oapd.inaf.it/cgi-bin/cmd}}} \citep{PARSEC} to constrain the stellar parameters theoretically through their evolution tracks, and also to model the combined stellar energy distribution (SED) of the four stars. To solve the inverse problem, the code employs a Markov Chain Monte Carlo (MCMC) based parameter search with an implementation of the generic Metropolis-Hastings algorithm \citep[see e.g.][]{ford05}. 

Our combined analysis is primarily based on the following observational inputs: $(i)$ the high-quality \textit{TESS} lightcurve (see Sect.~\ref{sec:TESSphot}), $(ii)$ the RV data available in the literature (see Sect.~\ref{subsect_RVdata}), $(iii)$ the ETV data calculated from all the available photometric observations (see Sect.~\ref{subsect:ETVdata} and Tables~\ref{tab:BGIndA_etv_TESS}--\ref{tab:BGIndA_etv_ground}) and, $(iv)$ the observed passband magnitudes of the target taken from standard catalogs (see Sect.~\ref{subsect:catalogdata} and Table~\ref{tab:catalogs}).

Note that, in its present form, {\sc Lightcurvefactory} is unable to handle period variations caused by non-few-body perturbations. Therefore, we did not model the small linear period variation of binary A which manifests itself in the form of quadratic deviations of the first few ETV points (see above, in Sect.~\ref{sect:classicETV}). Considering the fact that our analysis  depends primarily on the \textit{TESS} measurements obtained over the last 2.5 years, and on the radial velocity data gathered within a relatively narrow eight-year-long interval about a decade ago, we do not expect that such a small, long-term effect will have any significant influence on our results. Nevertheless, we will return to this question in Sect.~\ref{sec:discussion}. 

Regarding the archival photometric observations, we decided not to use the photometric fits to these lightcurves themselves for the analysis. This decision was based mainly on the fact that their large scatter was found to be of the same order as, or even higher than, the eclipse depths of the faint binary B. We did, however, utilize in our analysis the most relevant information that could be mined from these observations, namely the best of the mid-eclipse times that could be derived from these data.  Furthermore, the other benefit of these data is that they reveal that the eclipse depths of binary A have remained constant during the last $\sim40$~years, the relevance of which will be discussed later.

Before the analysis, we also took further preparatory steps on the \textit{TESS} lightcurve. In order to save computational time we binned the 2-min cadence data into 30-min bins, and in the following analyses we worked with the binned data. While carrying out some preliminary fitting runs on this 30-min binned Sector 1 lightcurve, we realized that the residual curve exhibits small amplitude, quasi-periodic variations. We had also found very similar patterns in the residual lightcurve at the end of the prior lightcurve disentangling process (Sect.~\ref{subsect:disentanglement}), i.e., after removing both binaries from the original time series.  We therefore concluded that these small quasi-cyclic variations cannot be the consequence of some misadjustements of the lightcurve parameters during our analysis, but should be real effects. In order to find the dominant frequencies of these fluctuations we calculated the power spectrum of the residual curve with a discrete Fourier-transform. We found two independent sets of frequency peaks. The frequencies of one set were close to the orbital frequency of binary A (and its multiples), while the other set was clearly related to the orbital frequency of binary B. 

Similar to what we have done in some of our previous work \citep[see e.g.][]{borkoetal18} we modelled these fluctuations during our analysis in the following manner.  We found that the use of the two most dominant frequencies of both sets of frequencies resulted in a significant improvement in the solutions. The process itself works as follows: in each trial step, after the removal of the blended EB lightcurves from the observed data, the residual lightcurve is modelled with harmonic functions of the four fixed frequencies, of which the eight (plus one) coefficients are obtained via matrix inversion. Then, this mathematical model of the residual lightcurve is added to the double binary model lightcurve and the actual $\chi^2$ value is calculated for this mixed model lightcurve. Finally note, since the fluctuations were found to be quasi-periodic instead of strictly periodic, we found that our process is the most effective if we use only a short section of the \textit{TESS} lightcurve. Therefore, for the main portion of our analysis we used only a seven-day-long section of the Sector~1 \textit{TESS} lightcurve.\footnote{We will show later, however, that by arbitrarily choosing another section of the lightcurve, we obtain very similar results, well within the $1\sigma$ statistical uncertainties of most of the adjusted parameters.}

The combined analyses were carried out in two different stages. In the first stage we worked only with astrophysical model-independent parameters. Therefore, we fitted simultaneously only the \textit{TESS} lightcurve, and the RV and ETV curves, but did not include SED data and theoretical stellar isochrones. During this phase the 20 adjusted parameters were as follows:
\begin{itemize}
\item[(i)] Seven lightcurve related parameters: the temperature ratios of $(T_2/T_1)_\mathrm{A,B}$ and $T_\mathrm{Ba}/T_{Aa}$, i.e. the secondary over primary temperature ratios of both binaries, and the ratio of the temperatures of the two primaries; the durations of the two primary eclipses $(\Delta t_\mathrm{pri})_\mathrm{A,B}$; the ratios of the radii in both pairs $(R_2/R_1)_\mathrm{A,B}$; and the gravity darkening coefficients of the two stars of binary A ($\beta_\mathrm{Aa,Ab}$). 
\item[(ii)] One parameter for each inner binary orbit, i.e. the observed inclinations $i_\mathrm{A,B}$ of the orbital planes of binary A and binary B, and five orbital parameters of the outer orbit: period ($P_\mathrm{out}$), time of periastron passage $\tau_\mathrm{out}$, eccentricity and argument of periastron $(e\cos\omega)_\mathrm{out}$ and $(e\sin\omega)_\mathrm{out}$, and the inclination $i_\mathrm{out}$.
\item[(iii)] Four mass-related parameters: the masses of the two primaries ($m_\mathrm{Aa,Ba}$), and the mass ratios of the two binaries ($q_\mathrm{A,B}$).
\end{itemize}

Regarding the other orbital parameters of the inner binaries, the periods ($P_\mathrm{A,B}$) of these EBs, as well as their orbital phase (through the time of an arbitrary primary eclipse -- $\mathcal{T}^\mathrm{pri}_\mathrm{A,B}$) were constrained internally through the ETV data. Furthermore, the eccentricities of both inner orbits were set to zero. Moreover, for the large $P_\mathrm{out}/P_\mathrm{A,B}$ ratios we found that all three orbits (two inner binary orbits and the outer orbit) can be considered as pure, unperturbed Keplerian motion. Due to this latter consideration our dataset does not contain any information about the positions of the orbital nodes relative to each other. Therefore, the sixth orbital element, the longitude of the node of each orbit ($\Omega_\mathrm{A,B;out}$) was fixed at zero. Finally, we note that the systemic radial velocity of the whole quadruple system ($\gamma$) was also constrained internally by minimizing the $\chi^2_\mathrm{RV}$ contribution a posteriori in each trial step.

Turning to the atmospheric parameters of the four stars, in contrast to our previous analyses, we now adjust the gravity darkening coefficients ($\beta$) of the strongly non-spheroidal components of the bright binary A. The reason is that, in contrast to the widely used classic model of \citet{lucy67} which predicts a unique gravity darkening coefficient of $\beta=0.32$ for all convective stars, recently \citet{claretbloemen11} have shown that the true relations are much more complicated.  This is especially true for stars close to the transition region between convective and radiative envelopes, where the components of binary A are located. On the other hand, in the case of binary B, we kept fixed the usual value of $\beta=0.32$ prescribed in Lucy's model. Other atmospheric parameters, such as the logarithmic limb-darkening coefficients $(x,y)_\mathrm{TESS}$ were interpolated in each trial step with the use of passband-dependent tables downloaded from the Phoebe 1.0 Legacy page\footnote{\url{http://phoebe-project.org/1.0/download}}. These tables are based on \citet{castellikurucz04} atmospheric models and are primarily used for the original version of the {\sc Phoebe} software \citep{Phoebe}. Furthermore, for the components of the bright binary A ($Aa$ and $Ab$), we include the reflection/irradiation effect into the lightcurve model and, therefore, we take into account the bolometric limb-darkening coefficients $(x,y)_\mathrm{bol}$, interpolating them in each trial step in the same manner as was done with the passband-dependent coefficients.

At this stage of the analysis we required only one further parameter that is undetermined by the model and, therefore, has to be set externally. This was the effective temperature of the primary of binary A, which was set (and kept fixed) according to the findings of \citet{rozyczkaetal11}.

At the end of this stage of the analysis we obtained accurate dynamical masses not only for the two members of the bright binary A but, in addition, we obtained the total dynamical mass of binary B.\footnote{This latter was driven mainly by the amplitude ratios of the cyclic ETV curves of the two binaries and also by the varying systemic RV of binary A (not to be confused with the systemic RV of the whole quadruple, $\gamma$, described above).}  Furthermore, the temperature ratio of the two primaries provides reliable information about the characteristics of the two stars forming the faint binary B. Finally at this stage, the orbital elements of the three orbits were also accurately determined.

In the next and final stage of the analysis, we included the SED information into the analysis  as well as the built-in \texttt{PARSEC} tables.  Now the seven lightcurve-related parameters described above were no longer adjusted but, instead, the radii and temperatures of all the four stars were constrained, i.e., recalculated at the beginning of each trial step by interpolating their values from the three-dimensional (mass, metallicity, age) grids of the \texttt{PARSEC} tables\footnote{The interpolation method was described in detail in \citet{borkoetal20}}.  During this phase of the analysis three additional adjustable quantities were introduced, including $(i)$ the metallicity ($[M/H]$) and $(ii)$ the (logarithmic) age of the quadruple. These two parameters, together with the mass of the given components, determined the position of each star within the \texttt{PARSEC} grids and, therefore, determine the interpolated fundamental stellar parameters and theoretical passband magnitudes.  The third parameter $(iii)$ was the stellar extinction $E(B-V)$ for the SED fitting. Moreover, while fitting the model SED to the dereddened observed SED points, the distance of the system comes in as an additional parameter. The software constrains this parameter a posteriori in each trial step by minimizing the value of $\chi^2_\mathrm{SED}$.

After some initial trials, however, we found it necessary to introduce a fourth, extra parameter to adjust, namely the age of the evolved component of binary A, in order to obtain model lightcurves which yield similarly low $\chi^2_\mathrm{LC}$ values to the ones obtained in the previous astrophysical model-independent stage. This procedure requires some further explanation. It is generally expected that the components of a close binary (multiple) system are coeval. Theories, however, allow for small departures from exact coevality \citep[see, e.g.][]{tokovinin18}, which during some critical rapid stages of stellar evolution might be significant. Furthermore, even in the case of exact coevality, the approximative nature of our interpolation method certainly carries with it inherent inaccuracies which might lead to modest discrepancies in the derived stellar parameters, especially during the very rapid sensitive evolutionary stage of the evolved star in binary A. Therefore, as a counterbalance to these uncertainties, we allowed for the age of the evolved component to be set independently from the other three stars. 

\begin{figure*}
\begin{center}
\includegraphics[width=0.49 \textwidth]{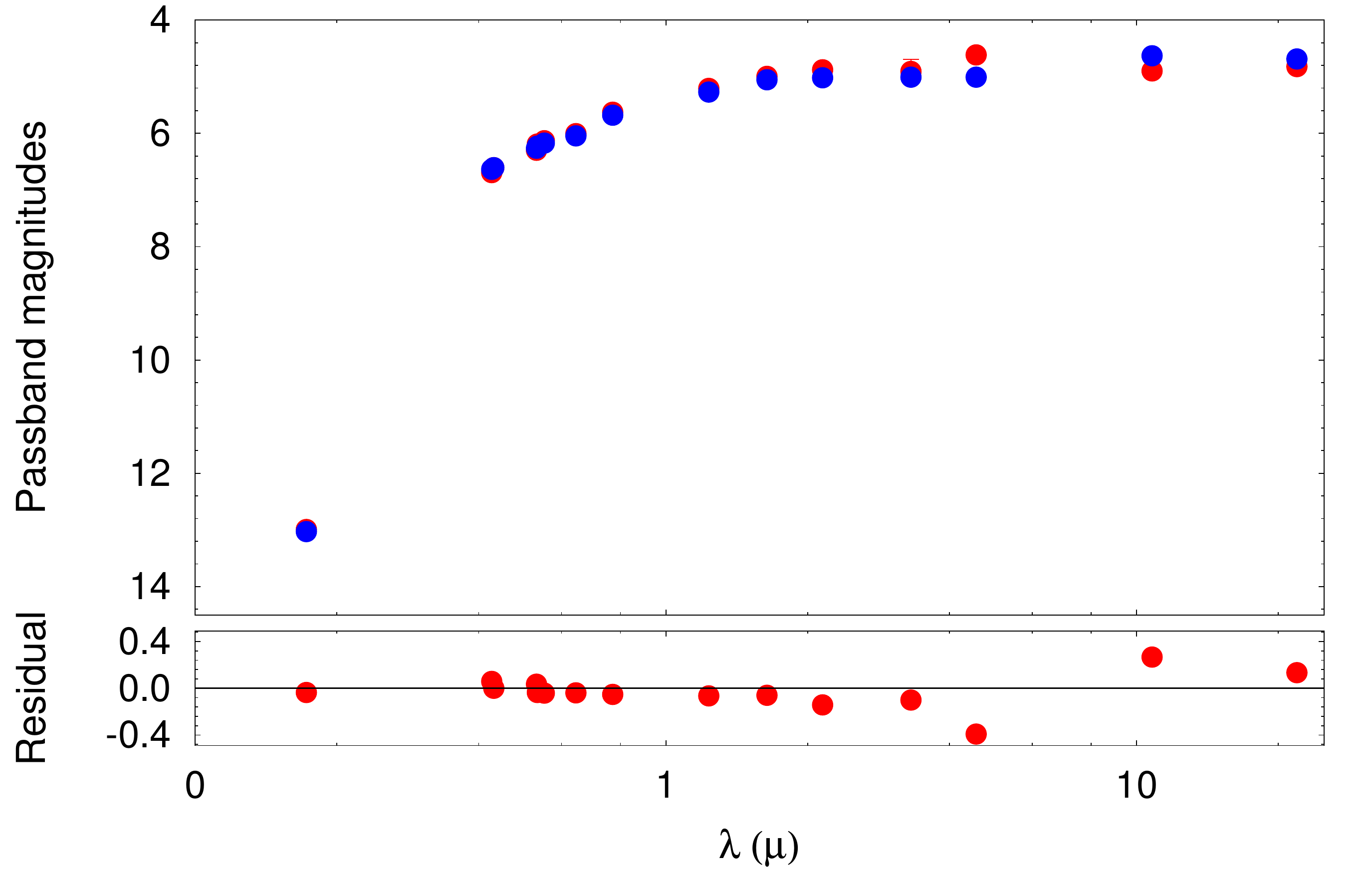} 
\includegraphics[width=0.49 \textwidth]{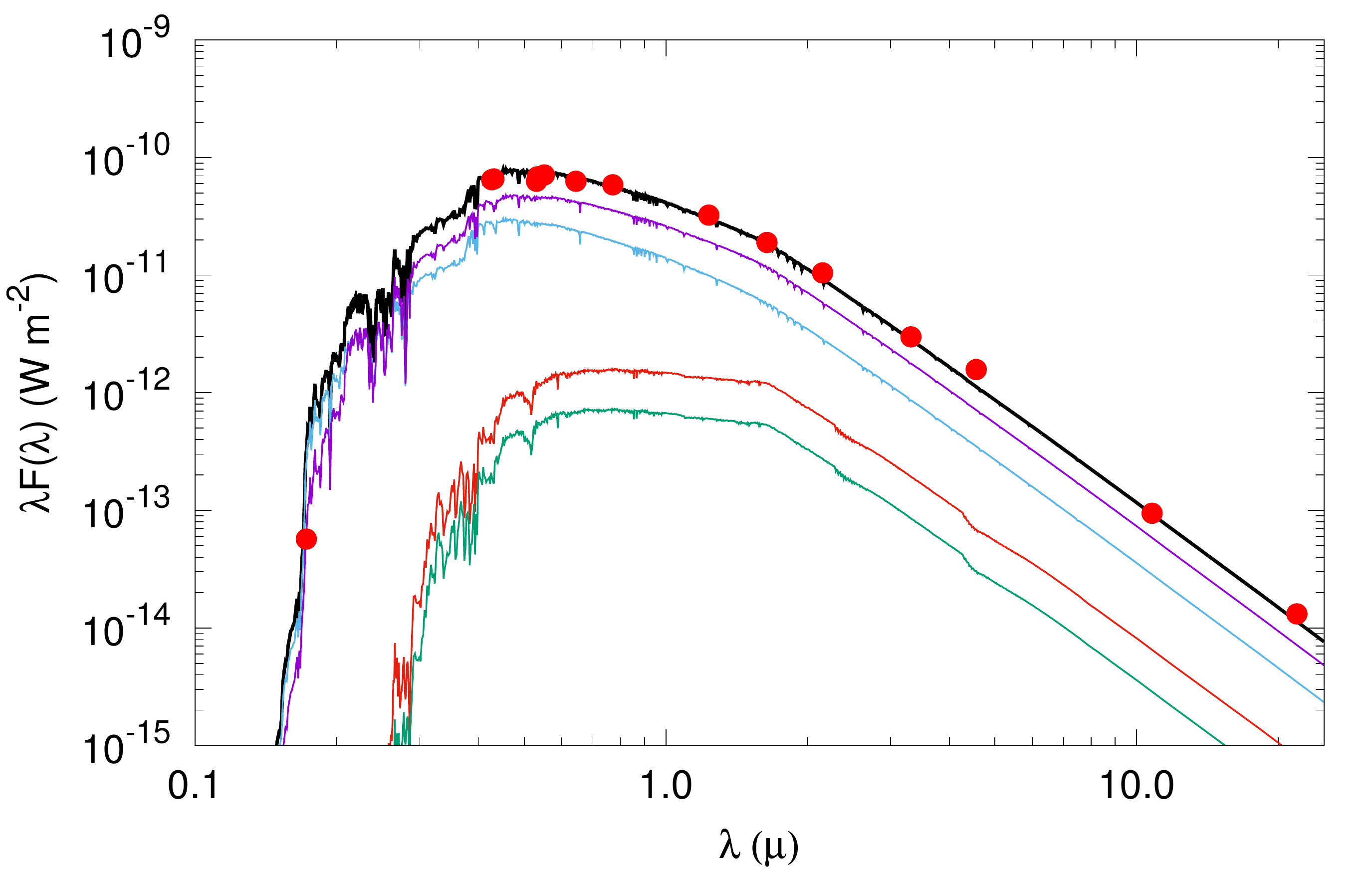}
\caption{The summed SED of the four stars of BG~Ind both in the magnitude and the flux domains. The {\it left panel} displays the cataloged values of the passband magnitudes (red filled circles; tabulated in Table~\ref{tab:catalogs}) versus the model passband magnitudes derived from the absolute passband magnitudes interpolated with the use of the \texttt{PARSEC} tables (blue filled circles).  In the {\it right panel} the dereddened observed magnitudes are converted into the flux domain (red filled circles), and overplotted with the quasi-continuous summed SED for the quadruple star system (thick black line). This SED is computed from the \citet{castellikurucz04} ATLAS9 stellar atmospheres models (\url{http://wwwuser.oats.inaf.it/castelli/grids/gridp00k2odfnew/fp00k2tab.html}). The separate SEDs of the four stars are also shown with thin green, black and purple lines, respectively.} 
\label{fig:sedfit} 
\end{center}
\end{figure*}

In Table~\ref{tab: syntheticfit} we tabulate the median values and the $1\sigma$ statistical uncertainties of the parameters obtained during the last stage of our analysis. The synthetic model light curves derived from the best-fit joint solution are displayed in Figs.~\ref{fig:lightcurve} and \ref{fig:wasp_lc}.  The corresponding ETV and RV curves are presented in Figs.~\ref{fig:etvs} and \ref{fig:RVAab_fold}, respectively. Finally, in the two panels of Fig.~\ref{fig:sedfit}, we illustrate the SED-fitting part of the combined solution both in the flux and the passband magnitude domain.

\begin{table*}
 \centering
\caption{Orbital and astrophysical parameters of BG~Ind from the joint photodynamical lightcurve, RV, ETV, SED and \texttt{PARSEC} isochrone solution.}
 \label{tab: syntheticfit}
\begin{tabular}{@{}llllll}
\hline
\multicolumn{6}{c}{orbital elements} \\
\hline
   & \multicolumn{3}{c}{subsystem}  \\
   & \multicolumn{2}{c}{A} & \multicolumn{2}{c}{B} & A--B \\
  \hline
  $P_\mathrm{a}$ [days] & \multicolumn{2}{c}{$1.464065_{-0.000002}^{+0.000002}$} & \multicolumn{2}{c}{$0.528349_{-0.000002}^{+0.000002}$} & $720.9_{-3.1}^{+3.4}$ \\
  $a$ [R$_\odot$] & \multicolumn{2}{c}{$7.602_{-0.043}^{+0.038}$} & \multicolumn{2}{c}{$3.025_{-0.016}^{+0.011}$} & $540.4_{-2.2}^{+2.7}$ \\
  $e$ & \multicolumn{2}{c}{$0$} & \multicolumn{2}{c}{$0$} & $0.209_{-0.048}^{+0.028}$ \\
  $\omega$ [deg]& \multicolumn{2}{c}{$-$} & \multicolumn{2}{c}{$-$} & $1.6_{-8.8}^{+9.2}$ \\ 
  $i$ [deg] & \multicolumn{2}{c}{$73.27_{-0.13}^{+0.06}$} & \multicolumn{2}{c}{$84.29_{-0.87}^{+0.85}$} & $85.5_{-6.3}^{+3.1}$ \\
  $\mathcal{T}^\mathrm{pri}$ [BJD - 2\,400\,000]& \multicolumn{2}{c}{$58326.1362_{-0.0012}^{+0.0011}$} & \multicolumn{2}{c}{$58325.8072_{-0.0022}^{+0.0025}$} &  \\
  $\tau$ [BJD - 2400000]& \multicolumn{2}{c}{$-$} & \multicolumn{2}{c}{$-$} & $58699_{-21}^{+14}$ \\
  \hline
  mass ratio $[q=m_\mathrm{sec}/m_\mathrm{pri}]$ & \multicolumn{2}{c}{$0.919_{-0.006}^{+0.010}$} & \multicolumn{2}{c}{$0.932_{-0.015}^{+0.014}$} & $0.483_{-0.005}^{+0.007}$ \\
  $K_\mathrm{pri}$ [km\,s$^{-1}$] & \multicolumn{2}{c}{$120.47_{-0.75}^{+1.12}$} & \multicolumn{2}{c}{$138.98_{-1.39}^{+1.23}$} & $12.57_{-0.24}^{+0.17}$\\ 
  $K_\mathrm{sec}$ [km\,s$^{-1}$] & \multicolumn{2}{c}{$130.99_{-0.50}^{+0.48}$} & \multicolumn{2}{c}{$149.25_{-1.13}^{+1.09}$} & $26.02_{-0.50}^{+0.31}$\\ 
  $\gamma$ [km\,s$^{-1}$] &  \multicolumn{4}{c}{$-$}  & $48.69_{-0.59}^{+0.29}$ \\
  \hline  
\multicolumn{6}{c}{stellar parameters} \\
\hline
   & Aa & Ab &  Ba & Bb & \\
  \hline
 \multicolumn{6}{c}{Relative quantities and atmospheric properties} \\
  \hline
 fractional radius$^b$ [$R/a$]  & $0.3084_{-0.0044}^{+0.0016}$  & $0.2096_{-0.0019}^{+0.0052}$ &  $0.2120_{-0.0014}^{+0.0013}$ & $0.2019_{-0.0020}^{+0.0018}$ \\
 fractional flux [in \textit{TESS}-band] & $0.6294_{-0.0153}^{+0.0064}$ & $0.3475_{-0.0060}^{+0.0165}$ & $0.0133_{-0.0008}^{+0.0009}$ & $0.0092_{-0.0007}^{+0.0008}$ \\
 $x_\mathrm{bol}^c$ & $0.676_{-0.001}^{+0.001}$ & $0.671_{-0.001}^{+0.001}$ & ... & ... \\
 $y_\mathrm{bol}^c$ & $0.174_{-0.003}^{+0.002}$ & $0.198_{-0.003}^{+0.002}$ & ... & ... \\
 $x_{TESS}^c$ & $0.631_{-0.001}^{+0.001}$ & $0.618_{-0.001}^{+0.001}$ & $0.720_{-0.001}^{+0.001}$ & $0.724_{-0.002}^{+0.001}$ \\
 $y_{TESS}^c$ & $0.347_{-0.002}^{+0.002}$ & $0.354_{-0.003}^{+0.002}$ & $0.281_{-0.005}^{+0.005}$ & $0.325_{-0.018}^{+0.019}$ \\
 $\beta^d$    & $0.11_{-0.07}^{+0.10}$    & $0.64_{-0.33}^{+0.29}$     & 0.32 & 0.32\\
 \hline
 \multicolumn{6}{c}{Physical Quantities} \\
  \hline 
 $m$ [M$_\odot$] & $1.432_{-0.024}^{+0.015}$ & $1.315_{-0.023}^{+0.026}$ & $0.688_{-0.011}^{+0.008}$ & $0.640_{-0.011}^{+0.010}$ \\
 $R^b$ [R$_\odot$] & $2.339_{-0.021}^{+0.016}$ & $1.592_{-0.019}^{+0.047}$ & $0.642_{-0.007}^{+0.005}$ & $0.611_{-0.009}^{+0.008}$ \\
 $T_\mathrm{eff}^b$ [K] & $6442_{-28}^{+29}$ & $6816_{-26}^{+26}$ & $4609_{-49}^{+48}$ & $4327_{-57}^{+62}$ \\
 $L_\mathrm{bol}^b$ [L$_\odot$] & $8.433_{-0.169}^{+0.199}$ & $4.934_{-0.179}^{+0.279}$ & $0.167_{-0.009}^{+0.009}$ & $0.118_{-0.009}^{+0.009}$ \\
 $M_\mathrm{bol}^b$ & $2.45_{-0.03}^{+0.02}$ & $3.04_{-0.06}^{+0.04}$ & $6.72_{-0.06}^{+0.06}$ & $7.09_{-0.08}^{+0.08}$ \\
 $M_V^b$            & $2.45_{-0.03}^{+0.03}$ & $3.02_{-0.06}^{+0.04}$ & $7.23_{-0.09}^{+0.10}$ & $7.83_{-0.13}^{+0.13}$ \\
 $\log g^b$ [dex]   & $3.852_{-0.005}^{+0.011}$& $4.150_{-0.016}^{+0.007}$ & $4.660_{-0.002}^{+0.003}$ & $4.672_{-0.004}^{+0.004}$ \\
 \hline
\multicolumn{6}{c}{Global Quantities} \\
\hline
$\log$(age)$^b,e$ [dex] & $9.40_{-0.01}^{+0.02}$ & \multicolumn{2}{c}{$9.32_{-0.01}^{+0.03}$} &\\
$[M/H]^b$  [dex]    &\multicolumn{4}{c}{$-0.189_{-0.037}^{+0.038}$} \\
$E(B-V)$ [mag]    &\multicolumn{4}{c}{$0.018_{-0.013}^{+0.015}$} \\
$(M_V)_\mathrm{tot}^b$           &\multicolumn{4}{c}{$1.93_{-0.03}^{+0.03}$} \\
distance [pc]                &\multicolumn{4}{c}{$69.7_{-0.9}^{+0.7}$}  \\  
\hline
\end{tabular}

\textit{Notes. }{$a$: Time of the inferior conjunction of the secondary component (i.e. mid-time of a primary eeclipse); $b$: Interpolated (or derived) from the \texttt{PARSEC} isochrones; $c$: interpolated linear ($x$) and logarithmic ($y$) limb-darkening coefficients. Note, bolometric coefficients used only during the calculation of the reflection effect, therefore they were not set for binary B; $d$: gravity darkening coefficients; $e$: the age of the evolved primary component of binary A was allowed to vary independently of the other three stars -- see text for details. }
\end{table*}

\section{Results and discussion}
\label{sec:discussion}

\subsection{Orbital configuration}

Our analysis confirms the hierarchical 2+2 type quadruple star nature of BG Ind. Thanks to the available high-quality \textit{TESS} photometry and the long-term ground-based photometric and spectroscopic observations, BG Ind now takes its place as (i) one of the most compact 2+2 quadruples known, as well as (ii) the quadruple system with the most accurately known stellar masses and other stellar parameters. The outer period of the system is found to be $P_\mathrm{out}=721\pm3$\,days, which is the shortest amongst doubly eclipsing quadruple systems with an accurately known outer period.\footnote{We emphasize that this holds only for {\em doubly eclipsing} 2+2 quadruples. The shortest period known 2+2 quadruple system, VW~LMi has a much shorter outer period of $P_\mathrm{out}=355$~days \citep{pribullaetal08,pribullaetal20}. Furthermore, in the case of the doubly eclipsing quadruple star EPIC\,220204960 \citet{rappaportetal17} reported that the outer period is very likely between 300 and 500~d, but an accurate value for that system is unknown.} Note, however, that despite the relatively short outer period, both outer to inner period ratios are large enough ($P_\mathrm{out}/P_\mathrm{A}\approx492$, $P_\mathrm{out}/P_\mathrm{B}\approx1365$) so that we do not expect readily measurable short-term mutual three- (four-) body perturbations.  In other words, all three orbits can be considered as essentially purely Keplerian. The outer orbit is moderately eccentric with $e_\mathrm{out}=0.21\pm0.05$, and is seen nearly along the direction of the minor axis ($\omega_\mathrm{out}=2\degr\pm9\degr$). 

These relatively small uncertainties in the BG~Ind quadruple system, however, should be treated with some caution. The two main reasons for this caveat can nicely be seen in the ETV plots in Figure~\ref{fig:etvs}. First, due to the unlucky fact that the outer period is nearly exactly equal to two years ($P_\mathrm{out}\approx1.973$\,yr), the annual observing seasons of the target can, and do, miss the most informative two parts of the ETV curve, i.e. its two extrema. Second, as was discussed above, the very first eight pre-WASP ETV points show clear deviations from the pure LTTE solution, and might indicate a continuous, constant increase in the orbital period of binary A, i.e., $\dot P_{\rm A}$. 

Though the modelling of $\dot P_{\rm A}$ was not included in the comprehensive spectro-photodynamical approach, its effect can be quantified by comparing the orbital parameters of the outer orbit obtained through the classic, analytic LTTE+quadratic solution of the ETV of binary A (Table~\ref{tab:ETV_lite}) with the detailed spectro-photodynamical model (Table~\ref{tab: syntheticfit}). As one can see, the outer period and eccentricity match well within their estimated uncertainties, while the argument of pericenter, the periastron passage time and the RV amplitudes are discrepant at the $2-3\sigma$ level. Therefore, we can conclude that, as was expected, the omission of the quadratic ETV term in the complex spectro-photodynamical analysis did not influence our basic solutions, but suggests that the actual uncertainties in the orbital elements should be somewhat larger than cited in Table~\ref{tab: syntheticfit}.

The inclination of the outer orbit is found to be $i_\mathrm{out}=86\degr\pm5\degr$. On the other hand, the inclinations of the two inner eclipsing binaries are found to be $i_\mathrm{A}=73\fdg1\pm0\fdg1$ and $i_\mathrm{B}=84\fdg3\pm0\fdg9$. From these values, and in the absence any information on the longitude of the nodes of the three orbits ($\Omega_\mathrm{A,B,out}$), the only thing one can say is that the whole quadruple system is certainly not perfectly flat. Since the mutual inclination of two planes cannot be smaller than the difference between the two observed inclinations of the planes considered (and, cannot be larger than their sum), the inclination of the bright binary A relative to the outer orbit must surely exceed $\approx13\pm5\degr$, but may even reach $90\degr$. (Similarly, the mutual inclination between the orbital plane of binary B and the outer orbital plane may be anywhere between coplanar and perpendicular.)  

As a consequence of such misalignments, one may expect the binary's orbital plane to precess. In that case, eclipse depth variations should be observed, or even the disappearance of the eclipses on a longer time-scale.  The period of forced precession of a binary orbital plane in a hierarchical triple system can be well approximated with the expression \citep[see e.g.][]{soderhjelm75}
\begin{equation}
(P_\mathrm{prec})_\mathrm{A}=\frac{4}{3}\frac{1+q_\mathrm{out}}{q_\mathrm{out}}\frac{P_\mathrm{out}^2}{P_\mathrm{A}}\left(1-e_\mathrm{out}^2\right)^{3/2}\left[\frac{C}{G_2}\cos(i_\mathrm{m})_\mathrm{A}\right]^{-1},
\end{equation}
where $C$ represents the total orbital angular momentum of the quadruple, while $G_2$ is the orbital angular momentum stored in the outer orbit. In the present situation it can be readily seen that $C/G_2\approx1$, i.e., the majority of the orbital angular momentum of the quadruple is stored in the outer orbit. Therefore, with $q_{\rm out} \equiv M_B/M_A \simeq 0.5$ one can easily show that $(P_\mathrm{prec})_\mathrm{A}\gtrsim4200$\,yr.  From this we conclude that there is no chance of detecting eclipse-depth variations given the available span of the observations.   

\subsection{Astrophysical properties and evolutionary status of the four stars}

Turning to the fundamental astrophysical parameters of the bright components of binary A, we compare our results to those of the former thorough analysis of \citet{rozyczkaetal11}. The masses of the stars found in the two analyses agree quite well: $1.432 \pm0.020$ vs.~$1.428\pm0.008$ for the more evolved component, and $1.315 \pm 0.025$ vs.~$1.293\pm0.008$ for its less evolved companion, where the first of each pair are from the current work.  This agreement is good to $\lesssim 1\,\sigma$, in units of our error bars.  The uncertainties given in \citet{rozyczkaetal11} are smaller than ours by  factors of 2-3. One should keep in mind, however, that \citet{rozyczkaetal11} estimated their uncertainties from an analysis of their own set of RV data which have smaller rms residuals than the combined set of RVs that we used. Furthermore, during their final RV analysis they corrected the RV values for the distortions of the stellar components with the use of the Wilson--Devinney code \citep{w-d71,wilson79}. In contrast to this, in our study, the effects of the distortions of the stars on the RV data are automatically taken into account within {\sc Lightcurvfactory}. Therefore, we consider our somewhat larger uncertainties to be more realistic.

The radii of the two stars in binary A exhibit slightly larger differences between the two studies.  Our analysis has yielded $R_\mathrm{Aa}=2.34\pm0.02\,\mathrm{R}_\odot$ for the evolved component, while \citet{rozyczkaetal11} obtained the somewhat smaller value of $2.29\pm0.02\,\mathrm{R}_\odot$.\footnote{Note, however, that they analysed six different lightcurves separately, and their results were scattered between $2.17\,\mathrm{R}_\odot$ and $2.40\,\mathrm{R}_\odot$.}  For the other less evolved star we found $R_\mathrm{Ab}=1.59\pm0.04\,\mathrm{R}_\odot$, in contrast to their somewhat larger value of $1.68\pm0.04\,\mathrm{R}_\odot$.  Keeping in mind, however, that the sum of the (fractional) radii of the two stars (i.e. $R_\mathrm{Aa}/a_\mathrm{A}+R_\mathrm{Ab}/a_\mathrm{A}$) is one of the most robustly determined parameters of an eclipsing binary's lightcurve (though, it does depend sensitively on the inclination), one can easily check, that this sum agrees well in the two solutions.\footnote{Note also that due to the significant tidal and rotational oblateness of the two stars, they will no longer be spherical; therefore, it should be clarified what is meant by `radius'. We cite the volume equivalent radius and assume that \citet{rozyczkaetal11} used the same definition. On the other hand, we note that in the above mentioned relation for the sum of the fractional radii, the so-called `side' radius, (i.e. measured in the star's equatorial plane, in the direction perpendicular to the line joining the two stars) should be the relevant one during eclipses. The volume equivalent and side radii for our stars, however, agree to better than 1\%.} 

The relatively small discrepancies in the individual radii between our analysis and that of \citet{rozyczkaetal11}, can likely be explained by some combination of the following three effects.  First, prior studies did not consider the small extra flux contribution (of $\approx2.2\%$) coming from binary B, and the lightcurve distortions caused by its eclipses. Second, we used the high quality \textit{TESS} photometry whose superiority over the former ground-based measurements is unquestionable. Third, we allowed for the reflection/irradiation effect which made our analysis more realistic, but this effect was not considered during the previous analyses. In conclusion, we emphasize again that the discrepancy in radii is fairly small.

We also found small departures in the effective temperatures of the two components of binary A compared with the previous results. Our results, which hinge to a large degree on the fit of the combined 4-star SED, resulted in slightly larger temperatures.  We found $T_\mathrm{Aa}=6442\pm29$\,K and $T_\mathrm{Ab}=6816\pm26$\,K  in contrast to $6350\pm260$\,K and $6650\pm230$\,K \citep{rozyczkaetal11}. Note. however, that our results are within the uncertainties of \citet{rozyczkaetal11}.  On the other hand, by using the temperatures given by \citet{rozyczkaetal11}, \citet{stassuntorres16} found a consistent SED solution for the binary. Of course, they did not consider the contribution of binary B, which might give a small excess at the red wing of the SED, and in turn which might force the fit toward slightly lower temperatures.

There is, however, an even more significant discrepancy between the system distance inferred from the SED solution and the trigonometric distance deduced from Gaia's measurements. Our solution has resulted in a photometric distance of $d=69.7\pm0.8$\,pc, while \citet{bailer-jonesetal18} using the Gaia DR2 measurements have obtained $d_\mathrm{DR2}=51.0\pm0.5$\,pc.\footnote{The parallaxes published in Gaia DR2 and EDR3 are well within $1\sigma$ of each other, and therefore we can assume that the distance derived from the EDR3 data will not differ significantly from the published DR2 distance.}  The situation is more complicated than this seemingly straightforward discrepancy. First, the trigonometric distance that can be calculated from the new reduction of HIPPARCOS parallaxes \citep{HIPrev} is $d_\mathrm{HIP}=67.1_{-2.2}^{+2.8}$, which is within $1\sigma$ of our result. Furthermore, \citet{stassuntorres16} used a similar SED modeling analysis to ours, and found a photometric distance of $d_\mathrm{Stassun+16}=66.7$\,pc. Again, this is much closer to our result and that of HIPPARCOS than to the Gaia distance.  However, since we know the fundamental stellar parameters of the dominant A binary quite accurately (including the bolometric luminosities) independent of the distance, and others have used partially different methods\footnote{\citet{rozyczkaetal11} have determined the temperatures with a combination of spectroscopic analysis and lightcurve fitting. \citet{stassuntorres16} utilized SED modelling.  And our study included a combination of lightcurve and SED fitting, where the consistency of the obtained stellar parameters were also probed by modelling the RV curves and identifying appropriate \texttt{PARSEC} isochrones tracks.} to find a very similar distance, we tentatively conclude that the published Gaia DR2 and EDR3 parallaxes are probably subject to some systematic error. This discrepancy might have arisen from the fact that the period of the outer orbit in BG Ind is very close to two years ($P_\mathrm{out}=1.973$\,yr) and, furthermore, the semi-major axis of BG~Ind~A's ellipse around the center of mass of the quadruple system is $a_\mathrm{out,A}\approx0.82$\,AU. Therefore, the combination of the orbital motion of the photocenter of binary A along the outer orbit and a period near 2 years, may be responsible for causing some problems with  Gaia's trigonometric parallax determination. 

Turning to the newly discovered binary B, we find it to be a pair of two mid K-type dwarfs with a near unit mass ratio of $q_\mathrm{B}=0.93\pm0.01$. Despite the fact that the flux contribution from this binary is only about 2\% in the \textit{TESS} photometric band, due to the high quality \textit{TESS} photometry of this rather bright quadruple, we were able to obtain quite good lightcurves (see Fig.~\ref{fig:ABfold}) and a robust dynamical model for binary B. The out-of-eclipse sections of the lightcurve of binary B are distorted, and we explain that by chromospheric activity, i.e. stellar spots, which are quite usual for stars with thick convective envelopes. As one can see in the middle panel of Fig.~\ref{fig:ABfold}, for Sector~1 data these variations can be well modelled (mathematically) with two Fourier-terms having frequencies equal to the orbital frequency and its first harmonic. On the other hand, in the case of Year 3 (i.e. Sectors~27 and 28 data) the same Fourier-representation was found to be less satisfactory, and we therefore added two additional harmonics of the binary B orbital frequency to the Fourier-representation (see the lowest panel of Fig.~\ref{fig:ABfold}).  However, even in this case, one can still notice some imperfections in the fit. We explain this fact with a possible rapid variation in the chromospheric activity which induces brightness fluctuations that cannot be well represented by a few smooth harmonics (even after averaging the lightcurve over a few weeks of the \textit{TESS} observations). Note, however, that this discrepancy is less than $\sim$100~ppm, and therefore, it would remain under the detection limit for any ground-based photometric observations.

The timings of the shallow eclipses from binary B are in accord with the 1.973-yr-periodicity in the ETV curve found from binary A, both of which are dominated by the light travel time effect (Fig.~\ref{fig:etvs}). Moreover, the dynamically deduced total mass of binary B, coupled with the dimensionless stellar parameters, lead to physical parameters of the stars in binary B.  And these, according to the \texttt{PARSEC} tables we used, are fully consistent with the parameters of two main-sequence K-dwarfs, having the same age and metallicity as those of the members of the bright binary A. Therefore, there is no question that the two eclipsing binaries form a compact, gravitationally bound, hierarchical quadruple star system.

Regarding the global parameters of the quadruple, the combined solution prefers a slightly metal-deficient abundance of $[M/H]=-0.19\pm0.04$ which, again, is in perfect agreement with the previous result of $[Fe/H]=-0.2\pm0.1$ by \citet{rozyczkaetal11}.  As mentioned before, we did not enforce strict coevality among the stellar components during our analyses and found an age of $\tau_\mathrm{Aa}=2.51\pm0.12$\,Gyr for the evolved primary, and $\tau_\mathrm{Ab,B}=2.14\pm0.10$\,Gyr for the three main-sequence components. These two ages differ by $\simeq 370 \pm 150$ Myr, or $\sim$7\% of the age, with a significance of only 2.5\,$\sigma$. We consider this discrepancy to be not a `small' departure from the coevality (though it does not have a high statistical significance). Our impression is that it might arise from the rapid rotation as well as the tidal distortion of the evolved star. Therefore, it is possible that the spherical stellar radius given by the \texttt{PARSEC} tables would not strictly equal the volume-equivalent radius of a strongly tidally distorted star.

Finally, we briefly discuss the question of the likely continuous period increase in the binary A period, detected through the systematic deviations of the very first ETV points from a simple linearly sloped LTTE model. Such period variations have been observed in a fair number of EBs. In the case of semi-detached and contact systems, the most common explanation is some kind of mass exchange between the stellar components. However, since all the previous studies have found that BG~Ind~A is a detached system, and our detailed analysis confirms this scenario, this period increase cannot be explained via mass transfer. (And this is not to mention the fact that, in this case, the increasing period would imply that the less evolved, lower mass secondary star would be the mass-donor, which is an unphysical scenario.) On the other hand, however, as was shown, e.g. by \citet{pringle85} and \citet{demircanetal06} mass loss from a close binary star (e.g., due to a stellar wind) always leads to an increasing period. Therefore, the quadratic ETV-term in BG~Ind~A might imply an enhanced stellar wind from the surface of the evolved component. For a quantitative study of this possibility, further observations of high-quality eclipse times over a longer time interval would be extremely useful. 

We also note that naturally, an LTTE effect driven by a more distant, low-mass fifth stellar component may also be the correct explanation. Obviously, the confirmation or refutation of this scenario also requires further eclipse follow up observations.

\section{Summary and Conclusions}
\label{Sect:Summary}

In this paper we report the discovery of the doubly eclipsing quadruple nature of the previously known, bright, southern eclipsing binary BG~Ind. We present the first comprehensive analysis of BG Ind in its entirety. {\em TESS} observations provided high-precision photometry covering two intervals of one and two months, respectively, and separated by two years.  Even though these high-quality {\em TESS}  observations covered only short segments of the outer orbit with $P_{\rm out} = 1.973$\,yr, we were able to use ground-based archival lightcurve and RV data to determine accurately the orbital and dynamical parameters of the system.

BG Ind is found to have one of the shortest outer periods among all quadruple systems having a 2+2 hierarchy. According to the recent version of the Multiple Stellar Catalog \citep{MSC} there are only five such systems (including BG~Ind) with outer periods shorter than 3 years. These are tabulated in Table~\ref{tbl:quadruples}.

The remarkably small number of such compact 2+2 quadruples that are known probably arises from an observational selection effect rather than for astrophysical reasons---specifically, they are quite difficult to discover. In contrast with the discovery of a third companion of a known binary star, which can be made by, e.g., astrometry, long-term RV measurements (or, in the case of an EB) ETV studies, or in exceptional cases, observing serendipitous extra eclipses, the binary nature of such a third component would remain hidden in most cases.  The only rare exceptions are when the second binary happens to be also an EB\footnote{On the other hand, despite the fact that such large photometric surveys such as, e.g., the gound-based Optical Gravitational Lensing Experiment \citep[OGLE;][]{udalskietal15}, or the  \textit{TESS} mission have observed hundreds of  lightcurves exhibiting blends of at least two EBs, the gravitationally bound nature of the blended EBs have been proven definitively for only a relatively small fraction of these objects \citep[see, e.~g.][]{zascheetal19}.} (as is the case in four of the five systems listed in Table~\ref{tbl:quadruples}), or it is bright enough to be observable as a second spectroscopic binary (as in the case of the fifth tabulated system, VW~LMi). Furthermore, another possibility for discovering the binarity of a component in a binary or multiple star system might arise from its faintness in comparison to its dynamically determined mass (as in the cases of $\kappa$~For, \citealp{tokovinin13} and $\zeta$~Cnc~C, \citealp{tokovinin17}).\footnote{In the case of the compact hierarchical triple system IU~Aur \citet{drechseletal94} and \citet{ozdemiretal03} have also concluded that the large third mass vs. small third light and weak spectroscopic signal discrepancies could be resolved by postulating that the third companion is a binary itself.  If this assumption was true, IU~Aur would be the shortest outer period 2+2 quadruple with $P_\mathrm{out}=294$\,d, but the system needs further investigations. Moreover, note that most recently \citet{marcadonetal20} have proposed that the $P_\mathrm{out}=180.4$ days period outer component of V1200~Cen might also be a binary, forming a more tight half-year-long period quadruple system.}

\begin{table}
\centering
\caption{List of hierarchical 2+2 quadruple systems with $P_\mathrm{out}<3$\,yr.}
\begin{tabular}{lllll}
\hline
Identifier & $P_\mathrm{out}$ & $P_\mathrm{A}$ & $P_\mathrm{B}$ & References \\
\hline
VW~LMi         & 355 &  0.478 & $7.931^*$ & 1 \\
EPIC~220204960 & 300--500 & 13.274 & 14.416    & 2 \\
BG~Ind         & 721 &  1.464 &  0.528    & 3 \\
TIC~278956474  & 858 &  5.488 &  5.674    & 4 \\
V994~Her       &1063 &  2.083 &  1.420    & 5 \\
\hline
\end{tabular}

{\em Notes.} $^*$ The $7.931$-day-period binary in VW~LMi does not exhibit eclipses.

{\bf References:} (1) \citet{pribullaetal08}; (2) \citet{rappaportetal17}; (3) present paper; (4) \citet{rowdenetal20}; (5) \citet{zascheuhlar16}
\label{tbl:quadruples} 
\end{table}

BG Ind is also one of only a very few compact quadruple systems where the key parameters of all fours stars are known with an accuracy of better than $\sim$3\%, including masses, radii, and $T_{\rm eff}$ values.  Likewise, the three sets of orbital parameters, such as periods, semi-major axes, eccentricities, and inclination angles are all known rather precisely. The one notable exception is that we do not have any way of determining the mutual inclination angles among the three orbital planes. The observational inclination angles are all close to edge on (i.e., 90$^\circ$), and we might surmise from statistical arguments that the most likely configuration is nearly coplanar for all three orbits. But we cannot be certain that this is indeed the case.

Future interferometric and astrometric observations may help to solve this problem. The semimajor axis of the outer orbit is $\simeq$ 36 mas, so in principle it is resolvable by speckle interferometry or adaptive optics.  However, the high contrast between the A and B binaries makes resolution with single-dish telescopes a challenging task.  A much better prospect is offered by long-baseline interferometers, e.g. the GRAVITY instrument at VLTI\footnote{\url{https://www.eso.org/sci/facilities/paranal/instruments/gravity.html}} 
\citep{Gravity}. The contrast in the K band is more favorable compared to the visible ($\Delta K=2.98$\,mag vs $\Delta V=4.73$\,mag), so the visibility and phase modulation caused by the outer pair can be well measured.  Furthermore, the spectral resolution of $R\sim4000$ offered by GRAVITY will allow detection of opposite phase shifts in the spectral lines of Aa and Ab at times near their maximum separation, thus enabling one to measure the orientation of the Aa,Ab orbit (its semimajor axis is 0.5\,mas) and, perhaps, even the orbit of Ba,Bb.

The orientation of the outer orbit on the sky will also be known from future Gaia data releases because the amplitude of the photocentric orbit is quite large, $\sim$12 mas. The proper motion anomaly (difference between the short-term proper motion measured by Gaia and the long-term proper motion deduced from the HIPPARCOS and Gaia positions) is quite large, ($+19.4$, $-13.7$)\,mas/yr \citep{brandt18}. Moreover, the long-term proper motion deduced from the HIPPARCOS and Gaia positions, ($+5.4$, $+29.0$)\,mas/yr, is close to ($+5.0$, $+30.2$)\,mas/yr measured by Gaia~EDR3 on a 2-yr time base that effectively averages the outer orbit, while Gaia~DR2 proper motion measured on a 1.5\,yr baseline is substantially different.

Finally since BG Ind is so bright, and the eclipses of binary A are relatively deep (at $\sim$15\%), we encourage amateurs to continue the eclipse timing.  The historical archival data are very helpful in this regard, but not as accurate as targeted observations of this star would be with small to modest sized telescopes. Furthermore, a secure verification of the suspected continuous period change of binary A also needs long-term, continuous follow up timing observations.

\section*{Acknowledgements}

The authors are grateful to Drs.~Keivan Stassun and Andrei Tokovinin for very valuable discussions.

T.\,B. acknowledges the financial support of the Hungarian National Research, Development and Innovation Office -- NKFIH Grant KH-130372.

P.\,M. is grateful for support from STFC research grant number ST/M001040/1.

We thank Allan R.~Schmitt for making his lightcurve examining software `{\sc LcTools}' freely available.

This paper includes data collected by the \textit{TESS} mission. Funding for the \textit{TESS} mission is provided by the NASA Science Mission directorate. Some of the data presented in this paper were obtained from the Mikulski Archive for Space Telescopes (MAST). STScI is operated by the Association of Universities for Research in Astronomy, Inc., under NASA contract NAS5-26555. Support for MAST for non-HST data is provided by the NASA Office of Space Science via grant NNX09AF08G and by other grants and contracts.

This work was partly based on data obtained from the ESO Science Archive Facility under request number 556483.

This work has made use of data  from the European  Space Agency (ESA)  mission {\it Gaia}\footnote{\url{https://www.cosmos.esa.int/gaia}},  processed  by  the {\it   Gaia}   Data   Processing   and  Analysis   Consortium   (DPAC)\footnote{\url{https://www.cosmos.esa.int/web/gaia/dpac/consortium}}.  Funding for the DPAC  has been provided  by national  institutions, in  particular the institutions participating in the {\it Gaia} Multilateral Agreement.

This publication makes use of data products from the Wide-field Infrared Survey Explorer, which is a joint project of the University of California, Los Angeles, and the Jet Propulsion Laboratory/California Institute of Technology, funded by the National Aeronautics and Space Administration. 

This publication makes use of data products from the Two Micron All Sky Survey, which is a joint project of the University of Massachusetts and the Infrared Processing and Analysis Center/California Institute of Technology, funded by the National Aeronautics and Space Administration and the National Science Foundation.

We  used the  Simbad  service  operated by  the  Centre des  Donn\'ees Stellaires (Strasbourg,  France) and the ESO  Science Archive Facility services (data  obtained under request number 396301).

\section*{Data availability}

The \textit{TESS} data underlying this article were accessed from MAST (Barbara A. Mikulski Archive for Space Telescopes) Portal (\url{https://mast.stsci.edu/portal/Mashup/Clients/Mast/Portal.html}). A part of the data were derived from sources in public domain as given in the respective footnotes. The derived data generated in this research and the code used for the photodynamical analysis will be shared on reasonable request to the corresponding author.



\bsp	
\label{lastpage}
\end{document}